\documentclass[preprint,review,11pt]{elsarticle}

\usepackage{graphicx}
\usepackage{amssymb}
\usepackage{float}
\usepackage{amsmath}
\usepackage[top=1cm]{geometry}
\usepackage{setspace}
\usepackage{lineno}
\usepackage{hyperref}
\hypersetup{
    colorlinks,
    citecolor=blue,
    linkcolor=blue,
    backref=true
}
\usepackage{multirow}
\usepackage{adjustbox}
\usepackage{caption}
\usepackage[font={small,bf}]{caption}
\usepackage{tabularx,booktabs}

\usepackage{algorithm}
\usepackage{algpseudocode}
\algdef{SE}{Begin}{End}{\textbf{begin}}{\textbf{end}}
\algnewcommand{\parState}[1]{\State%
    \parbox[t]{\dimexpr\linewidth-\algmargin}{\strut\hangindent=\algorithmicindent \hangafter=1 #1\strut}}
    
\algnewcommand\algorithmicswitch{\textbf{switch}}
\algnewcommand\algorithmiccase{\textbf{case}}
\algnewcommand\algorithmicassert{\texttt{assert}}
\algnewcommand\algorithmicother{\textbf{otherwise}}
\algnewcommand\Assert[1]{\State \algorithmicassert(#1)}%
\algdef{SE}[SWITCH]{Switch}{EndSwitch}[1]{\algorithmicswitch\ #1\ \algorithmicdo}{\algorithmicend\ \algorithmicswitch}%
\algdef{SE}[CASE]{Case}{EndCase}[1]{\algorithmiccase\ #1}{\algorithmicend\ \algorithmiccase}%
\algdef{SE}[OTHER]{Other}{EndOther}[1]{\algorithmicother\ #1}{\algorithmicend\ \algorithmicother}%
\algtext*{EndSwitch}%
\algtext*{EndCase}%
\algtext*{EndOther}%

\usepackage{calrsfs}
\DeclareMathAlphabet{\pazocal}{OMS}{zplm}{m}{n}

\journal{International Journal of Production Economics}
\newgeometry{left=2cm,right=2cm,top=2cm,bottom=2cm}

\begin{document}
%\SetCommentSty{mycommfont}
%\newcommand\mycommfont[1]{\small\ttfamily\textcolor{blue}{#1}}
\begin{frontmatter}


\title{Reverse Logistics Network Design to Estimate the Economic and Environmental Impacts of Take-back Legislation: A Case Study for E-waste Management System in Washington State}

\author[mymainaddress]{Hadi Moheb-Alizadeh\corref{mycorrespondingauthor}}
\cortext[mycorrespondingauthor]{Corresponding author}
\ead{hadi.mohebalizadeh@gmail.com}

\author[mysecondaryaddress]{Amir Hossein Sadeghi}

\author[mysecondaryaddress2]{Amirreza Sahebi fakhrabad}

\author[mysecondaryaddress3]{Megan Kramer Jaunich}

\author[mysecondaryaddress4]{Eda Kemahlioglu-Ziya}

\author[mysecondaryaddress5]{Robert B Handfield}

\address[mymainaddress]{North Carolina State University, Raleigh, NC USA; email: \href{mailto:hadi.mohebalizadeh@gmail.com}{hadi.mohebalizadeh@gmail.com} }

\address[mysecondaryaddress]{North Carolina State University, Raleigh, NC USA; email: \href{mailto:asadegh3@ncsu.edu}{asadegh3@ncsu.edu}}

\address[mysecondaryaddress2]{North Carolina State University, Raleigh, NC USA; email: \href{mailto:asahebi@ncsu.edu}{asahebi@ncsu.edu}}

\address[mysecondaryaddress3]{North Carolina State University, Raleigh, NC USA; email: \href{mailto:mkjaunic@ncsu.edu}{mkjaunic@ncsu.edu}}

\address[mysecondaryaddress4]{North Carolina State University, Raleigh, NC USA; email: \href{mailto:ekemahl@ncsu.edu}{ekemahl@ncsu.edu}}

\address[mysecondaryaddress5]{North Carolina State University, Raleigh, NC USA; email: \href{mailto:rbhandfi@ncsu.edu}{rbhandfi@ncsu.edu}}

\begin{abstract}
In recent years, recycling and disposal of end-of-life (EOL) electronic products has attracted considerable attention in response to concerns over resource recovery and environmental impacts of electronic waste (e-waste). In many countries, legislation to make manufacturers responsible for taking e-waste at the end of their useful lives either has been adopted or is being considered. In this paper, by capturing different stages in the life-cycle of EOL electronic products (or, e-waste) generated from private or small-entity users, we develop two different formulations of a reverse logistics network, i.e. system-optimum model and user-optimum model, to estimate both economic and environmental effects of take-back legislation. In this system, e-waste is collected through user drop-off at designated collection sites. While we study the whole reverse logistics network associated with recycling and remanufacturing of e-waste in the system-optimum model and obtain an optimum solution from the policy maker's perspective, we split the logistics network into two distinct parts in the user-optimum model in order to derive an optimum solution from the users' standpoint. Implementing the proposed models on an illustrative example shows how they are capable of estimating the economic and environmental impacts of take-back legislation in various stages of e-waste's life-cycle. %Case study and Insights ...
\end{abstract}


\end{frontmatter}

%\linenumbers

\section{Introduction}
Sending products to the landfill continues to be the predominant disposal route for many products at the end of their useful life. A number of environmental challenges arise in this context: a) landfill capacity is limited, b) disposition to landfill and incineration can produce unwanted environmental emissions, and c) sending whole products to landfill means the residual value in the materials and components that make up the product are lost. Recently, companies are pushed by legislators to consider the environmental and social impacts of their supply chain networks. \cite{jaunich2020life} presented a holistic framework for analyzing electronic waste management systems and their supply chain network under two scenarios of re-selling and recycling based on energy cost, employee wage, and facility development costs. 

Take-back legislation has been introduced as a policy tool to divert used products from landfills. The legislation holds manufacturers responsible for collecting and properly disposing of their products at the end of usage periods. In this paper, we develop two generic reverse supply chain models to explore both the economical and environmental impacts of take-back legislation. Across the world, there exist examples of such legislation enacted for different types of products, e.g., Waste Electric and Electronic Equipment (WEEE) Directive in the European Union, the Japanese Specified Home Appliances Recycling Law (SHARL), and the Japanese PC Recycling System \citep{boks1998international}. Although the US is the second largest producer of e-waste (6.9 million tons in 2019) in the world \citep{noauthor_staggering_2022}, there is no federal law in the United States mandating the take-back or recycling of e-waste to the date \citep{schumacher2021waste}. Despite the fact that state-level initiatives are crucial for the management of e-waste, the reality that only 25 states have rules in place and they are all different makes recycling in the US extremely difficult \citep{schumacher2019towards, schumacher2021waste}. The COVID-19 pandemic and sever chip shortage has revealed the importance of recycling the electronic devices at the same time (** needs reference **). 

The majority of take-back legislation establishes a collection target, such that a manufacturer is (financially and/or physically) responsible to collect a minimum fraction of the products it's sold in the previous period. In addition, manufacturers must ensure that the collected end-of-use items are treated in an environmentally sound way, and most manufacturers meet this requirement by ensuring products are properly recycled by facilities with the appropriate certification (e.g., R2, e-Stewards). Another alternative used is the reuse of components or whole products. While diverting potentially dangerous substances from landfills is an environmental benefit, the true economical and environmental impacts of take-back legislation are studied in few studies. The belief held by policymakers and environmentalists alike is that take-back legislation is universally beneficial for environment, such as enhancements to the cascade separation and purification process's chemical use efficiency and clean energy use efficiency in lithium batteries \citep{zhang2021evaluation}, or saving time, money, reducing waste, and polluting less in recycling construction wastes \citep{cyril2021potential}. However, \cite{esenduran2016take} raises concerns about the validity of these beliefs by suggesting that actions taken by companies in response to legislation may increase the total environmental impact of a product. In addition, measurement systems for establishing the relative benefits and costs of take-back legislation are lacking. The majority of the work on take-back legislation in the operations management literature \citep{atasu2009efficient, esenduran2016take} involves high-level economic modeling, which fails to capture the different stages in the life-cycle of the product in detail. In computing the total environmental impact, researchers \citep{moazzeni2022dynamic, reddy2020effect} use data from existing life-cycle analysis studies, but some key information, such as the energy requirements of the remanufacturing stage, is missing from the model, and ad hoc estimates are used instead. Furthermore, the life-cycle analysis studies that these papers refer to were not conducted for understanding the environmental impact of legislation, and hence, do not fully capture key aspects of take-back such as reverse logistics or remanufacturing operations. Therefore, a fresh approach is required to establish a methodology for the estimation of life-cycle costs associated with take-back legislation decisions in order to frame government legislation and manufacturers' product disposition policies alike.

By developing two different mathematical programming models, we propose a framework that can be used to estimate the environmental and economic impacts of take-back legislation as applied to the representative e-waste management system. The framework is based on life-cycle process models that estimate cost and emissions parameters per unit of e-waste managed at each node. These parameters are used by the optimization model to identify the optimal paths for the set of devices to meet the objective in consideration of the system constraints. Rather than using disparate data coming from different sources as was done in earlier research, this model was developed alongside a data set that is focused on a set of products most commonly covered by legislation in the United States.

Our research will try to answer questions that are important for both manufacturers facing legislation as well as policymakers, including:  
\begin{enumerate}
\itemsep=0pt
\item If take-back legislation can indeed increase environmental impact as suggested by earlier research, which stage (or stages) of the life-cycle make the greatest contribution?
\item Can the environmental impact of legislation be (further) reduced through supply chain optimization (e.g., optimally locating collection points, optimizing transportation routes, or increasing the efficiency of recycling facilities)?
\item Can a cost model be developed and applied to identify the environmentally optimal treatment option for different types of products covered under the legislation?
\item From an environmental regulatory policy perspective, is the cost of implementing legislation justified by the reduction in environmental impact for all products covered under the legislation?
\end{enumerate}  

\section{Literature Review}
Our paper is related to the literatures on reverse logistics design, waste management, and waste management policy (i.e., take-back legislation). In this section, we provide an overview of the work most relevant to ours. 
\subsection{Reverse Logistics (RL)}
Due to the increasing environmental concerns, resource reduction, and depleting landfill capacities in many countries, optimizing reverse logistics operations for returned and end-of-life products receive growing attention in the last decades \citep{rocha2021life}. Effective implementation requires the establishment of suitable logistical systems for product flow from consumers to producers \citep{fleischmann2000characterisation}. The design of product recovery networks is one of the important and challenging issues in RL. In a RL network, used products originate from multiple sources and are transported to various types of recovery facilities. Typically, these networks are represented as mixed integer programs and the models are solved either to optimality using commercial solvers or through heuristic approaches. Examples of papers following this approach include \citep{fleischmann2000characterisation, shih2001reverse, jayaraman2003design, min2006genetic}. The physical location of facilities and transportation links need to be chosen to convey used products from their former users to a producer and to future markets again \citep{fleischmann2000characterisation}. developed a model for optimizing the collection and recycling processes of end-of-life (EOL) computers and home appliances in Taiwan. \cite{hu2002reverse} proposed a cost-minimization model for a multi-time-step, multi-type hazardous-waste RL system. They presented application cases to demonstrate the feasibility of their proposed approach. \cite{kucsakci2019optimization} presented a stochastic programming-based approach by which a deterministic location model for product recovery network design may be extended to account explicitly for uncertainties. They applied it to a representative real case study on recycling hazardous products, reusing products, and full recycling in Turkey. \cite{min2006genetic} determined the number and location of centralized return centers using a nonlinear mixed-integer programming model and a genetic algorithm that solved the RL problem involving product returns. \cite{li2022robust} developed a location-allocation model to determine the number and locations of the waste disposal plants in the region of Guangzhou (China) while \cite{bautista2006modeling} focused on selecting the locations of municipal waste collection points in Barcelona.\cite{kannan2010genetic} presented a development of genetic algorithm (GA) model for recycling spent batteries.\cite{gharibi2021mixed} developed a mixed-integer linear programming (MILP) model to maximize the profit of a reverse logistics network and presented a case of mobile phones and digital camera remanufacturing. \cite{sasikumar2010multi} developed a mixed integer nonlinear programming (MINLP) model to maximize the profit of reverse logistics network and presented a case of truck tire remanufacturing. \cite{zhao2016improved} developed a multi-objectives MILP model for network design for regional hazardous waste classified by certain properties, including  treatment technology, industrial distribution, and hazardous characteristics. \cite{mutha2009strategic} proposed a mathematical model for the design of a reverse logistics network that consisted of retailers, warehouses, reprocessing centers, remanufacturing factories, distribution and recycling centers, spare markets, disposal sites, and suppliers. They considered modular product structures with different disposal and recycling fractions for each module of each product in the model. \cite{santana2021refurbishing} analyzed the feasibility of a sustainable RL adopted by phone manufacturing industries, their results show that the refurbishing process stops waste disposal of 4.5 tons of Electrical and electronic waste (EEW) in Brazil. \cite{yu2020reverse} proposed a multi-period multi-objective MILP for a medical waste reverse logistics network to responsively address the rapid growth of medical waste within the planning window during the COVID-19 in keeping with reducing the risk of epidemic spread.

In addition to the studies provided in this section, readers are referred to \cite{prajapati2019bequeath} for a comprehensive survey on reverse logistics network design models.

\subsection{Waste management system (WMS)}
\cite{antunes1999location} studied the solid WMS of Central Portugal and develops a MIP model, combining elements of a p-median model and of a capacitated facility location model with transshipment, in which a maximum distance between sources and transfer stations, and between transfer stations and landfills is imposed. \cite{mitropoulos2009exact} studied a three-echelon network design problem with central treatment facilities, transfer stations and landfills, where the goal is to minimize the total cost of the solid WMS. \cite{ghiani2012capacitated} studied the problem of minimizing the total number of collection sites to be located in a SWM system, chosen among a set of candidate locations. Such an objective ensured not only the reduction of the impact due to the presence of the collection bins close to the residential sites, but also the reduction of the overall cost related to the collection phase. \cite{levis2013generalized} proposed a life-cycle optimization framework for municipal solid waste management. Their developed framework was built on a mixed-integer programming problem of the waste management supply chain and was capable of identifying the optimal waste management infrastructure under a variety of objective functions and constraints. \cite{santibanez2017dynamic} illustrated a three-stage MINLP-based approach helping companies in Mexico to locate plant capacities and material flows. \cite{anwar2018optimization} evaluated the location planning of a centralized WMS in rural areas of developing countries to optimize material and energy recovery costs. The configuration in this study includes technologies such as reuse (or recycling), landfilling, refuse-derived fuel, and composting. \cite{tong2021understanding} presented a system dynamics model for municipal solid WMS in Vietnam by proposing a model aiming at analyzing the contribution and activities of the informal sector, focusing on its roles and impacts. \cite{salehi2022designing} presented a
novel real-time WMS in smart cities based on the Internet of things aiming at minimizing pollution, and optimization of value recovery with considering the uncertainty of waste flow in separation centers. \cite{shaban2022optimization} presented a model for municipal solid WMS in Fayoum Governorate (Egypt) by proposing a MILP model aiming at minimizing the net daily cost, determining the best location for collection sites among a given set of candidate locations, and optimal flow of waste of the system. 

\subsection{Take-back legislation}
In this category, \cite{atasu2009efficient} holistically identified efficiency conditions for extended producer responsibility types of legislation and discusses the economic and environmental impacts while characterizing the right policy for take-back. \cite{walther2008negotiation} develop a decentralized coordination mechanism for allocation of WEEE and choice of disassembly levels to independent WEEE recycling companies. \cite{grunow2009designing} developed an approach providing an efficient assignment for all involved actors in WEEE program (e.g. the producers, the municipalities, and the collective schemes) by presenting an optimization-based decision support tool for the coordinating government institution in Denmark. \cite{ding2020production} formulated an optimization model for optimal production and carbon emission reduction (CER) rate decisions. They analyzed how carbon taxes and take-back legislation affect firms' production and CER decisions. \cite{li2021managing} presented a single-period stylized model for manufacturers to maximize their profit, considering mandated collection and recycling target rates for production, and recycling strategies. Their study shows that strict mandated collection legislation may not always have a negative impact on profits. However, more stringent mandated legislation may bring undesirable outcomes, because the recycling rate is not affected by extreme recycling benefits. \cite{nagurney2005reverse} developed a multitier e-cycling network model with a variational inequality formulation in a Cournot game. The formulation provides endogenous equilibrium prices and material flows between tiers. \cite{hammond2007closed} extended the work of \cite{nagurney2005reverse} and proposed a closed-loop supply chain model for WEEE in a Cournot pricing game with perfect information. They numerically found that minimum recovery targets stimulate manufacturers’ reverse chain activities, which contradicts later findings by \cite{li2021managing}, regarding disincentivization of manufacturers to implement green product design whose maximum recycling benefits are either too low or too high.

\section{System description}
The reverse logistics network of a typical waste management system comprises sources of returned products, drop-off sites, primary processors to dismantle the returned products and separate into reusable components, constituent materials and residual waste, and secondary processors to remanufacture the recovered materials or dispose of residual, which are configured as illustrated in Figure 1. The decisions to be made in the associated reverse logistics network include the number, location and capacity of the facilities involved, aggregate quantities and material flows of e-waste product through the network. 

In a representative waste management system, individuals and organizations bring or arrange for waste products to be taken to designated drop-off sites, where devices are sorted and resold if functional, or stored until they are shipped to a primary processing facility. Among all available collection options, fixed drop-off locations are used as the collection points in this paper because such systems have been the most extensively used and found to be the most cost-effective \citep{renckens2015basel, fan2018modeling}. An appropriate collection site can be selected by taking into consideration the geographic location, the ease and convenience to consumers, and the population distribution \citep{moheb2021efficient}. Transportation is also an important issue to address in a waste management system. With permanent collection sites, residents are responsible for the transportation to the collection site, while the transportation of collected waste products to the primary processing facilities is the responsibility of the processors \citep{kang2005electronic, patil2020comprehensive}.

Once transported to the primary processing facility, the e-waste might be divided into two categories: reusable or recyclable. The equipment and parts that can be reused are sorted for resale purposes and everything else is recycled through either a manual dismantling process, or one with some level of automation. The extent to which dismantling is automated is dependent on the technology and business model of each primary processing facility. This study models manual, semi-automated, and automated facilities.

After sorting recoverable constituent materials of e-waste products at a primary processor, each individual recovered material stream is sent to the appropriate secondary processing or remanufacturing facility. Reprocessed items including metals, plastics, glass, etc. are output from the system to be sold as commodities. The rate of material recovery at a given processor will depend on various parameters such as the size of the facility and the product under consideration \citep{kang2005electronic,ding2019recovery}. These materials offset the production of approximately equivalent virgin materials with respect to energy use and environmental emissions, as appropriate. Residual waste (e.g., cardboard, non-recyclable wood) may be sent to a landfill or incinerator.

\section{Problem Definition and Model Formulation} 
\label{sec: model}
In order to estimate the impact of take-back legislation on the performance of the reverse supply chain network depicted in Figure \ref{fig:sc}, we develop two models called system-optimum versus user-optimum. The system optimum model is developed from the policy makers’ perspective; it optimizes the supply chain performance over the whole network in a centralized fashion (curb to grave). On the other hand, from the end user standpoint, the user-optimum model consists of two models to separately optimize distinct parts of the network (namely, the drop-off network and the recycling network) in a decentralized manner.

The associated indices, parameters and decision variables used to build the proposed models are presented in Appendix.

\subsection{System-Optimum Model}
Policy makers aim to design public systems that operate cost-effectively to perform a necessary function. This translates to a perspective in which optimizing the entire system with respect to some criteria defined over all components of the system is a logical goal, if a reasonable representation of the system under investigation could be modeled. In other words, the policy makers prefer to have a system with all components operating optimally. To meet such a requirement, we develop a mixed integer linear programming (MILP) model that optimizes the total cost incurred and total emission generated through the e-waste reverse logistics network illustrated in Figure \ref{fig:RL network}. Under the system-optimum model, all the components in the network are committed to execute the optimal decision determined by the policy maker. Recall that, in this system, residents drop off e-waste items at one of the open drop-off sites using their personal vehicles. E-waste items are then transported from the respective drop-off sites to one of the available primary processors. Afterward, recovered materials and waste streams are sent from each primary processor to the appropriate secondary processing facilities or recovered material re-processing facilities.

\subsubsection{Objective Functions}

\textit{Cost objective function:} The total cost incurred in the logistics network associated with a waste management system includes four main components: transportation cost, processing cost, fixed cost and revenue from sale of recovered materials. In other words, the total cost equals the summation of transportation, processing and fixed costs minus reselling revenue. 

The transportation cost through the logistics network is derived as equation \eqref{eq:TrCo}, where $ty$ is the average number of trips per participating household per year to return e-waste products to any drop-off sites, and $df_c$ is a parameter to indicate the fraction of each trip to drop-off site $c$ at city $t$ in county $u$ that is dedicated to that purpose (default value is one). Furthermore, $pp_h/hs$ gives the number of households inhabiting in residence area $h$. Multiplying this number by the participation rate, the average number of trips per participating household and dedicated fraction yields to the total number of trips by households in area $h$ to drop-off site $c$ at city $t$ in county $u$.
\begin{flalign} \label{eq:TrCo}
TrCo = & \sum_{i=1}^{I}\sum_{h=1}^{H}\sum_{c=1}^{C} \dfrac{pp_h}{hs} \cdot pt \cdot ty \cdot df_c \cdot tco_{ihc}^{res-drp} d_{hc}^{res-drp} RTD_{ihc} +  \\& \nonumber
 \sum_{i=1}^{I}\sum_{c=1}^{C}\sum_{p=1}^{P} tco_{icp}^{drp-pri} d_{cp}^{drp-pri} DTP_{icp} + \nonumber
\sum_{j=1}^{J}\sum_{p=1}^{P}\sum_{s=1}^{S} tco_{jps}^{pri-sec} d_{ps}^{pri-sec} PTS_{jps} &&
\end{flalign}

Moreover, equations \eqref{eq:PrCo} and \eqref{eq:FiCo} characterize the processing and fixed costs, respectively:
\begin{flalign} \label{eq:PrCo}
PrCo =  & \sum_{i=1}^{I}\sum_{h=1}^{H}\sum_{c=1}^{C} (1-re_i^{drp}) r_{ih} pco_{ic}^{drp}  RTD_{ihc}+ \sum_{i=1}^{I}\sum_{c=1}^{C}\sum_{p=1}^{P} (1-re_i^{pri}) pco_{ip}^{pri} DTP_{icp} +  \\ &  \nonumber
\sum_{j=1}^{J}\sum_{p=1}^{P}\sum_{s=1}^{S} (1-re_i^{dec}) pco_{js}^{sec} PTS_{jps} && 
\end{flalign}
\begin{flalign} \label{eq:FiCo}
FiCo = & \sum_{c=1}^{C} fcc_c X_c  + \sum_{p=1}^{P} fcp_p Y_p  + \sum_{s=1}^{S} fcs_s R_s  &&
\end{flalign}

On the hand, the total revenue gained by selling resalable products in all kinds of facilities is derived in equation \eqref{eq:SeRe}:
\begin{flalign} \label{eq:SeRe}
SeRe = & \sum_{i=1}^{I}\sum_{h=1}^{H}\sum_{c=1}^{C} re_i^{drp} r_{ih} pcr_{ic}^{drp}  RTD_{ihc} - \sum_{i=1}^{I}\sum_{c=1}^{C}\sum_{p=1}^{P} re_i^{pri} pcr_{ic}^{pri} DTP_{icp} - \\ & \nonumber
\sum_{j=1}^{J}\sum_{p=1}^{P}\sum_{s=1}^{S} re_j^{sec} pcr_{js}^{sec} PTS_{jps} &&
\end{flalign}

Hence, the objective function minimizing the total cost incurred is defined as follows:
\begin{flalign}
Min \quad TC_{system } = TrCo + PrCo + FiCo - SeRe &&  
\end{flalign}

\textit{Emission objective function:} In the same way, the objective function minimizing total emission generated in the logistics network is characterized as follows:
Hence, the objective function minimizing the total cost incurred is defined as follows:
\begin{flalign}
Min \quad EM_{system } = TrEm + PrEm - SeEm &&  
\end{flalign}
where $TrEm$, $PrEm$ and $SeEm$ denote transportation emission, processing emission and reselling emission offset, respectively, derived as \eqref{eq:TrEm}, \eqref{eq:PrEm}, and \eqref{eq:SeEm}:
\begin{flalign} \label{eq:TrEm}
TrEm =  &\sum_{i=1}^{I}\sum_{h=1}^{H}\sum_{c=1}^{C} \dfrac{pp_h}{hs} \cdot pt \cdot ty \cdot df_c \cdot env_{ihc}^{res-drp} d_{hc}^{res-drp} RTD_{ihc}+ \\&\nonumber
         \sum_{i=1}^{I}\sum_{c=1}^{C}\sum_{p=1}^{P} env_{icp}^{drp-pri} d_{cp}^{drp-pri} DTP_{icp} +  \sum_{j=1}^{J}\sum_{p=1}^{P}\sum_{s=1}^{S} env_{jps}^{pri-sec} d_{ps}^{pri-sec} PTS_{jps} &&
\end{flalign}
\begin{flalign} \label{eq:PrEm}
PrEm = & \sum_{i=1}^{I}\sum_{h=1}^{H}\sum_{c=1}^{C} (1-re_i^{drp}) r_{ih} env_{ic}^{drp}  RTD_{ihc} + 
 \sum_{i=1}^{I}\sum_{c=1}^{C}\sum_{p=1}^{P} (1-re_i^{pri}) env_{ic}^{pri} DTP_{icp} + \\& \nonumber
 \sum_{j=1}^{J}\sum_{p=1}^{P}\sum_{s=1}^{S} (1-re_i^{dec}) env_{js}^{sec} PTS_{jps} &&
\end{flalign}
\begin{flalign} \label{eq:SeEm}
SeEm = & \sum_{i=1}^{I}\sum_{h=1}^{H}\sum_{c=1}^{C} re_i^{drp} r_{ih} ecr_{ic}^{drp}  RTD_{ihc} + 
 \sum_{i=1}^{I}\sum_{c=1}^{C}\sum_{p=1}^{P} re_i^{pri} ecr_{ic}^{pri} DTP_{icp} +  \\& \nonumber 
 \sum_{j=1}^{J}\sum_{p=1}^{P}\sum_{s=1}^{S} re_j^{sec} ecr_{js}^{sec} PTS_{jps} &&
\end{flalign}

\subsubsection{Constraints}
 The constraints included in the system optimum model are categorized into six types: flow balance constraints, capacity constraints, minimum shipment constraints, constraints on number of opened facilities, legislation constraints and decision variables constraints.
 
 \textit{Flow balance constraints:} Constraint \eqref{eq:FeBaC1} implies that the summation of all fractions of wasted items shipped from each residence area to all drop-off sites should be naturally equal to one. It implies that no e-waste product is remained in residence areas. Moreover, after reselling the resalable products, the rest of wastes are transported from each drop-off site to primary processors. Hence, given a drop-off site $c$ in city $t$ at county $u$, constraint \eqref{eq:FeBaC2} requires that the summation of all remained e-waste products coming from all residence areas must be equal to the sum of all wastes sent from that drop-off site to primary processors. Such a requirement needs also to be met for primary processors, i.e. after processing remained products, all output materials in each primary processor should be transported to secondary processors, denoted by constraint \eqref{eq:FeBaC3}. 
\begin{flalign}
        \label{eq:FeBaC1} & \sum_{c=1}^{C} RTD_{ihc} = 1; \hspace{0.5cm} \forall i,h \\ 
        \label{eq:FeBaC2} & \sum_{p=1}^{P} DTP{icp}  - \sum_{h=1}^{H} r_{ih} RTD_{ihc}(1-re_i^{drp})= 0; \hspace{0.5cm} \forall i,c \\
        \label{eq:FeBaC3} & \sum_{s=1}^{S} PTS{jps}  - \sum_{i=1}^{I}\sum_{c=1}^{C} q_{ji} eff_{jp}^{pri} DTP_{icp}(1-re_i^{pri})= 0; \hspace{0.5cm} \forall j,p &&
\end{flalign}
 
 \textit{Capacity constraints:} Once a drop-off site is opened on a candidate location, the total amount of incoming e-waste products from residence areas should be less than or equal to its corresponding capacity. Moreover, once a drop-off site is not going to be located on a potential location, no wastes should be transported to it. Constraint \eqref{eq:CapC1} implies the later two conditions. In addition, the same requirements are satisfied by constraints \eqref{eq:CapC2} and \eqref{eq:CapC3} for primary processors and secondary processors, respectively, i.e. they require that all processors should operate under their corresponding capacities and any idle processors are required to be closed. 
\begin{flalign}
        \label{eq:CapC1}  & \sum_{h=1}^{H} r_{ih} RTD_{ihc} \leq X_c cap_{ic}^{drp}; \quad \forall i,c \\
        \label{eq:CapC2}  & \sum_{c=1}^{C} DTP_{icp} \leq Y_p cap_{ip}^{pri}; \quad \forall i,p \\
        \label{eq:CapC3}  & \sum_{p=1}^{P} PTS_{jps} \leq R_s cap_{js}^{sec}; \quad \forall j,s &&
\end{flalign}

 \textit{Minimum shipment constraints:} This kind of constraints implies that a minimum amount of e-waste products/materials should be transported to a facility in order for it to operate in practice. Moreover, they prevent an opened facility to be idle. In this regard, constraints \eqref{eq:ShipC1}-\eqref{eq:ShipC3} satisfy minimum shipment requirements for drop-off sites, primary processors and secondary processors, respectively.
 \begin{flalign}
         \label{eq:ShipC1}  & \sum_{h=1}^{H} r_{ih} RTD_{ihc} \geq X_c lim_{ic}^{drp}; \quad \forall i,c \\
         \label{eq:ShipC2}  & \sum_{c=1}^{C} DTP_{icp} \geq Y_p lim_{ip}^{pri}; \quad \forall i,p \\
         \label{eq:ShipC3}  & \sum_{p=1}^{P} PTS_{jps} \geq R_s lim_{js}^{sec}; \quad \forall j,s && 
\end{flalign}

 \textit{Constraints on number of opened facilities:} Using this type of constraints, we set a lower bound on the number of opened facilities of each type. Constraints \eqref{eq:NoFaC1}-\eqref{eq:NoFaC3} imply that the number of opened drop-off sites, primary processors and secondary processors should be greater than or equal to their corresponding minimum required numbers, respectively.
 \begin{flalign}
        \label{eq:NoFaC1}  & \sum_{c=1}^{C} X_c \geq nof^{drp} \\
        \label{eq:NoFaC2}  & \sum_{p=1}^{P} Y_p \geq nof^{pri} \\
        \label{eq:NoFaC3}  & \sum_{s=1}^{S} R_s \geq nof^{sec}  && 
\end{flalign}

 \textit{Decision variables constraints:} Constraint \eqref{eq:DecC1} requires the fraction of e-waste shipped from residence areas to drop-off sites lies between zero and one. That all physical flows are positive continuous quantities is satisfied by constraint \eqref{eq:DecC2}. Moreover, constraint \eqref{eq:DecC3} defines $X_c$, $Y_m$, $Z_a$, $T_s$ and $R_e$ as binary decision variables.
\begin{flalign}
        \label{eq:DecC1} & RTD_{ihc} \in [0,1] \hspace{0.5cm} \forall i,h,c \\
        \label{eq:DecC2} & DTP_{icp}, PTS_{jps} \geq 0 \hspace{0.5cm} \forall i,j,c,p,s \\
        \label{eq:DecC3} & X_c, Y_p, R_s \in \{0,1\} \hspace{0.5cm} \forall c,p,s &&
\end{flalign}

\subsection{User-Optimum Level} \label{sec:userOptimum}
The system optimum model developed in the previous section provides optimal device and material throughputs for each node in the network for the defined system in terms of the selected objective (i.e., least cost, least emissions). The least environmental impact solution may be optimal from the policy maker standpoint, whereas from the perspective of manufacturers paying for an EPR program, the least cost solution is likely to be the preferred solution. 

Regardless of the objective function used, in order to minimize total cost or emission of the network in the system optimum model, the residents of an area are assigned to a drop-off site to deliver e-waste products. However, there is no guarantee that the residents necessarily follow what the policy-maker enforces in practice. A resident may decide to go to the closest drop-off site in reality regardless how much the total cost and emission of the reverse logistics network are impacted by such a decision. However, the drop-off sites that are the closest to residence areas may not be assigned by the model when minimizing total cost and emission through the entire logistics network. In order to address this situation, a user-optimum model is designed in this section, which consists of the following two separate programming models:
\begin{itemize}
    \item Model (User, I) -- In this model, the residence areas are assigned to the closest drop-off sites to deliver their e-waste products.
    \item Model (User, II) -- The collected e-waste products in drop-off sites from Model I are transported to the primary and secondary facilities for further processing.
\end{itemize}

Clearly, the optimal solutions obtained from these two separate models together form a sub-optimal solution to the system optimum model. 

\subsubsection{Model (User, I)}
\textit{Objective function}

\textit{Cost objective function}: The cost objective function of Model I consists of transportation cost from residence areas to drop-off sites as follows:

\begin{flalign} \label{eq:Co1}
Min \quad TC_{(user,I)} = & \sum_{i=1}^{I}\sum_{h=1}^{H}\sum_{c=1}^{C} \dfrac{pp_h}{hs} \cdot pt \cdot ty \cdot df_c \cdot tco_{ihc}^{res-drp} d_{hc}^{res-drp} RTD_{ihc} &&
\end{flalign}

\textit{Emission objective function}: The objective function minimizing emission in Model I is defined as $Min \quad Em_{(user,I)} = TrEm_{(user,I)}$ where $TrEm_{(user,I)}$ denotes the transportation emission derived as: 

\begin{flalign} \label{eq:Em1}
Min \quad  TrEm_{(user,I)} = & \sum_{i=1}^{I}\sum_{h=1}^{H}\sum_{c=1}^{C} \dfrac{pp_h}{hs} \cdot pt \cdot ty \cdot df_c \cdot env_{ihc}^{res-drp} d_{hc}^{res-drp} RTD_{ihc} &&
\end{flalign}

\textit{Constraints}:

The constraints of Model I are also categorized into flow balance constraints, capacity constraints, minimum shipment constraints, Constraints on number of opened facilities, legislation constraints and decision variables constraints.

\textit{Flow balance constraints}: All products in residence areas should be transported to drop-off sites. Hence, the sole flow balance constraint in Model I is defined as constraint \eqref{eq:FeBaC1} in system optimum model.

\textit{Capacity constraints}: The amount of wasted items shipped to a drop-off site should be less than or equal to its respective capacity. Therefore, the capacity constraint of Model I is the same as constraint \eqref{eq:CapC1} in system optimum model.

\textit{Minimum shipment constraints}: The minimum shipment constraint of Model I is characterized as constraint \eqref{eq:ShipC1} in system optimum model.

\textit{Constraints on number of opened facilities}: This type of constraints in Model I is the same as constraint \eqref{eq:NoFaC1} in system optimum model.

\textit{Policy/Legislation constraints}: Any policy/legislation constraint associated with transporting products from residence areas to drop-off sites are enforced here. 

\textit{Decision variables constraints}: Two decision variables of Model I include $RTD_{ihctu}$  and $X_{ctu}$. Their feasible ranges are determined by constraints \eqref{eq:DecC1}-\eqref{eq:DecC3} in baseline scenario, respectively.

\subsubsection{Model (User, II)}

As mentioned before, the collected e-waste in drop-off sites are then transported to primary and secondary processors. Hence, excluding the cost and emission components associated with transferring e-waste to drop-off sites leads to objective functions and constraints of Model II. 

\textit{Objective functions}

\textit{Cost objective function}: The total cost incurred in Model II includes four principal components: transportation cost, processing cost, fixed cost and reselling revenue. These components are respectively defined as follows:

\begin{flalign}
\label{eq:TrCo2}
TrCo_{(user,II)} = & \sum_{i=1}^{I}\sum_{c=1}^{C}\sum_{p=1}^{P} tco_{icp}^{drp-pri} d_{cp}^{drp-pri} DTP_{icp} + 
\sum_{j=1}^{J}\sum_{p=1}^{P}\sum_{s=1}^{S} tco_{jps}^{pri-sec} d_{ps}^{pri-sec} PTS_{jps} \\
\label{eq:PrCo2}
PrCo_{(user,II)} =  &  \sum_{i=1}^{I}\sum_{c=1}^{C}\sum_{p=1}^{P} (1-re_i^{pri}) pco_{ip}^{pri} DTP_{icp} +
\sum_{j=1}^{J}\sum_{p=1}^{P}\sum_{s=1}^{S} (1-re_i^{dec}) pco_{js}^{sec} PTS_{jps} \\
\label{eq:FiCo2}
FiCo_{(user,II)} = & \sum_{p=1}^{P} fcp_p Y_p  + \sum_{s=1}^{S} fcs_s R_s \\
\label{eq:SeRe2}
SeRe_{(user,II)} = & \sum_{i=1}^{I}\sum_{c=1}^{C}\sum_{p=1}^{P} re_i^{pri} pcr_{ic}^{pri} DTP_{icp} + 
\sum_{j=1}^{J}\sum_{p=1}^{P}\sum_{s=1}^{S} re_j^{sec} pcr_{js}^{sec} PTS_{jps} &&
\end{flalign}

Hence, the cost-minimizing objective function of Model II is defined as
\begin{flalign} \label{eq:Co2}
Min \quad TC_{(user,II)} = TrCo_{(user,II)} + PrCo_{(user,II)} + FiCo_{(user,II)} - SeRe_{(user,II)} &&
\end{flalign}

\textit{Emission objective function}: Likewise the components of total cost minimizing objective function derived above, the objective function minimizing total emission in Model I corresponds transportation emission, processing emission and reselling emission offset. These components are characterized as follows: 

\begin{flalign}
\label{eq:TrEm2}
TrEm_{(user,II)} =  & \sum_{i=1}^{I}\sum_{c=1}^{C}\sum_{p=1}^{P} env_{icp}^{drp-pri} d_{cp}^{drp-pri} DTP_{icp} +  \sum_{j=1}^{J}\sum_{p=1}^{P}\sum_{s=1}^{S} env_{jps}^{pri-sec} d_{ps}^{pri-sec} PTS_{jps} \\
\label{eq:PrEm2}
PrEm_{(user,II)} = & \sum_{i=1}^{I}\sum_{c=1}^{C}\sum_{p=1}^{P} (1-re_i^{pri}) env_{ic}^{pri} DTP_{icp} + 
 \sum_{j=1}^{J}\sum_{p=1}^{P}\sum_{s=1}^{S} (1-re_i^{dec}) env_{js}^{sec} PTS_{jps} \\
\label{eq:SeEm2}
SeEm_{(user,II)} = & \sum_{i=1}^{I}\sum_{c=1}^{C}\sum_{p=1}^{P} re_i^{pri} ecr_{ic}^{pri} DTP_{icp} + 
\sum_{j=1}^{J}\sum_{p=1}^{P}\sum_{s=1}^{S} re_j^{sec} ecr_{js}^{sec} PTS_{jps} &&
\end{flalign}

In this case, the total emission is derived as:
\begin{flalign} \label{eq:Em2}
Min \quad Em_{(user,II)} = TrEm_{(user,II)} + PrEm_{(user,II)} - SeEm_{(user,II)} &&
\end{flalign}

In is worth mentioning that the total cost and emission through the whole of the reverse supply chain under user optimum model are derived as:

\begin{flalign} \label{eq:TCUse}
TC_{user} = & min\{TC_{(user,I)}\} + min\{TC_{(user,II)}\}+ \\ &\nonumber  \sum_{c=1}^{C} fcc_c X_c + \sum_{i=1}^{I}\sum_{h=1}^{H}\sum_{c=1}^{C} (1-re_i^{drp}) r_{ih} pco_{ic}^{drp}  RTD_{ihc} - \sum_{i=1}^{I}\sum_{h=1}^{H}\sum_{c=1}^{C} re_i^{drp} r_{ih} pcr_{ic}^{drp}  RTD_{ihc} &&
\end{flalign}

\begin{flalign} \label{eq:EmUse}
Em_{user} = & min\{Em_{(user,I)}\} + min\{Em_{(user,II)}\}+ \\ &\nonumber  \sum_{i=1}^{I}\sum_{h=1}^{H}\sum_{c=1}^{C} (1-re_i^{drp}) r_{ih} env_{ic}^{drp}  RTD_{ihc}  - \sum_{i=1}^{I}\sum_{h=1}^{H}\sum_{c=1}^{C} re_i^{drp} r_{ih} ecr_{ic}^{drp}  RTD_{ihc} &&
\end{flalign}

\textit{Constraints}

In Model II, there are also six types of constraints including flow balance constraints, capacity constraints, minimum shipment constraints, constraints on number of opened facilities, legislation constraints and decision variables constraints.

\textit{Flow balance constraints}: Recall that the e-waste collected in each drop-off site in Model I should now be transported to the rest parts of the logistics network. On the other hand, all wastes in residence areas should be transported to drop-off sites. Hence, the flow balance constraints for drop-off sites are derived as follows:

\begin{flalign} \label{eq:FeBaCUse1}
\sum_{p=1}^{P} DTP_{icp} - (1-re_i^{drp}) rq_{ic} = 0; \quad \forall i, c &&
\end{flalign}

The flow balance constraint associated with primary processors is the same as constraints \eqref{eq:FeBaC3} in system optimum model. 

\textit{Capacity constraints}: All capacity constraints in Model II are related to primary and secondary processors and are the exactly same as constraints \eqref{eq:CapC2} and \eqref{eq:CapC3} in system optimum model. 

\textit{Minimum shipment constraints}: This type of constraints in Model II is characterized constraints \eqref{eq:ShipC2} and \eqref{eq:ShipC3} in system optimum model.

\textit{Constraints on number of opened facilities}: All constraints associated with minimum number of opened facilities in Model II are the same as constraints \eqref{eq:NoFaC2} and \eqref{eq:NoFaC3} in system optimum model.

\textit{Policy/Legislation constraints}: Any policy/legislation constraint associated with transporting products from residence areas to drop-off sites are enforced here. 

\textit{Decision variables constraints}: Excluding the decision variable $RTD_{ihctu}$ and $X_{ctu}$ that is related to model I, constraints \eqref{eq:DecC2} and \eqref{eq:DecC3} in system optimum model determine the feasible ranges of all decision variables in Model II.

\section{Solution Method}
The following steps outline the solution procedure for the model described in the previous section. First, the multiple objective functions are combined into a single objective function using the $\epsilon$-constraint method. Secondly, the resulting single objective programming model is solved using robust optimization, a technique that addresses uncertainty in the problem parameters and their solutions.

\subsection{Epsilon Constraint}
The multi-objective problem programming model presented in section~\ref{sec:userOptimum} aims to find a balance between economic, environmental, and social sustainability. To address this multi-objective problem, the $\epsilon$-constraint method is chosen for its simplicity and wide use in similar problems. This method has been successfully applied in various multi-objective scenarios in the past \citep{fakhrzad2018green,guillen2010global,moheb2019sustainable}.

The proposed user model aims to minimize both total cost and $CO_2$ emissions and can be summarized as follows:
\begin{flalign} \label{eq:multi-objective}\nonumber 
Min_{} \quad & \{TC_{user}, Em_{user}\} \\
s.t. \quad & x \in S &&
\end{flalign}

The proposed user model involves minimizing user total cost ($TC_{user}$) and $CO_2$ emissions ($Em_{user}$), represented by the decision variables in the vector $x$, within the feasible solution space ($S$). Using the $\epsilon$-constraint method, the multi-objective problem is transformed into a single objective programming model by selecting one of the objective functions as the primary objective and expressing the remaining objectives as constraints with defined bounds \citep{miettinen2012nonlinear}. If the total user cost objective~\eqref{eq:TCUse} is chosen as the primary objective, the resulting single objective programming model includes the total $CO_2$ emission objectives as the constraint.

\begin{flalign} \label{eq:epsilon-method}\nonumber 
Min_{} \quad & TC_{user} \\ \nonumber
s.t. \quad & Em_{user} \leq  Em_{user}^{min} + v \Delta \epsilon_{Em} \\ 
x \in S &&
\end{flalign}

Where $v = 0,1,...,V$ and $\Delta \epsilon_{Em} = \frac{Em_{user}^{max} - Em_{user}^{min}}{V}$.

To determine the minimum values for the total $CO_2$ emission and objective function, the following steps are followed:
\begin{enumerate}
    \item Find the optimal solution for each objective function in $S$. Then, create a set containing the optimal solutions for the $TC_{user}$, and $Em_{user}$ objective functions, represented as $X_{TC}^{*}$, and $X_{Em}^{*}$ respectively, we call the solution space $\Re = \{X_{TC}^{*}, X_{Em}^{*} \}$.
    \item Find the values of objective function $TC_{user}$ for $X_{Em}^{*}$, and $Em_{user}$ for $X_{TC}^{*}$.
    \item Find the minimum and maximum value of $Em_{user}$ as $Em_{user}^{min} = min\{Em_{user}(x), x \in \Re\}$ and $Em_{user}^{max} = max\{Em_{user}(x), x \in \Re\}$
\end{enumerate}

This paper employs the modification suggested by \cite{mavrotas2009effective} to ensure that the optimal solution of equation~\eqref{eq:epsilon-method} is also a Pareto optimal solution for the original multi-objective problem outlined in equation~\eqref{eq:multi-objective}. This is achieved by converting the constraints associated with the added objective functions into equalities by introducing slack variable ($S$) and then incorporating them as penalties in the single objective function. As a result, equation~\eqref{eq:epsilon-method} is transformed into the following model:

\begin{flalign} \label{eq:eepsilon-TCUse}\nonumber 
Min_{} \quad & TC_{user}^\prime =  min\{TC_{(user,I)}\} + min\{TC_{(user,II)}\} + \\ &\nonumber 
\sum_{c=1}^{C} fcc_c X_c + \sum_{i=1}^{I}\sum_{h=1}^{H}\sum_{c=1}^{C} (1-re_i^{drp}) r_{ih} pco_{ic}^{drp}  RTD_{ihc} - \\ & \sum_{i=1}^{I}\sum_{h=1}^{H}\sum_{c=1}^{C} re_i^{drp} r_{ih} pcr_{ic}^{drp} RTD_{ihc} + \theta S &&
\end{flalign}
$s.t.$ \quad \eqref{eq:FeBaC1}-\eqref{eq:DecC3}
\begin{flalign} \label{eq:epsilon-EmUse} \nonumber
\quad & min\{Em_{(user,I)}\} + min\{Em_{(user,II)}\}+  \\ \nonumber 
& \sum_{i=1}^{I}\sum_{h=1}^{H}\sum_{c=1}^{C} (1-re_i^{drp}) r_{ih} env_{ic}^{drp}  RTD_{ihc}  - \\ 
& \sum_{i=1}^{I}\sum_{h=1}^{H}\sum_{c=1}^{C} re_i^{drp} r_{ih} ecr_{ic}^{drp}  RTD_{ihc}  + S =  Em_{user}^{min} + v \Delta \epsilon_{Em}  \\
& X_c \in \{0,1\}, RTD_{ihc},S \geq 0; \qquad \forall i,h,c &&&&
\end{flalign}

The value of $\theta$ is set to a sufficiently small number (typically between $10^{-3} $ and $10^{-6}$) that does not impact the objective function. By solving the current single-objective programming model with a particular value of $v$, a Pareto optimal solution for the original multi-objective problem~\eqref{eq:multi-objective} is obtained. This process is repeated for multiple values of $v$, resulting in a set of $v+1$ Pareto optimal solutions that defines the Pareto front.

\subsection{An overview on robust optimization}
In this section, an overview of the robust optimization approach proposed by Bertsimas and Sim (2004) is presented. To do so, the following linear programming model is considered:
\begin{flalign} \nonumber 
Min_{} \quad &\sum_{j}^{} c_{j}x_{j}\\ \nonumber 
s.t. \quad &\sum_{j}^{} \Tilde{a}_{ij}x_{j} \leq b_{i}; \qquad\forall i\\
&x_{j}\geq 0; \qquad\forall j &&
\end{flalign}
where the technological coefficients $\Tilde{a}_{ij}$ are assumed to be uncertain. In other words, each coefficient $\Tilde{a}_{ij}$ is regarded as an independent, symmetric and bounded parameter, which can take values in $[a_{ij} - \hat{a}_{ij},a_{ij} + \hat{a}_{ij}]$, i.e. $\Tilde{a}_{ij} \in [a_{ij} - \hat{a}_{ij},a_{ij} + \hat{a}_{ij}]$. In this definition, $a_{ij}$ and $\hat{a}_{ij}$ denote the nominal value and the maximum deviation from the nominal value, respectively. Associated with each row \textit{i} in problem (1) is $J_{i}$, which is defined as the set of all coefficients in row \textit{i} that are subject to uncertainty. Furthermore, a scaled deviation $\eta_{ij} \in [-1,1]$ is defined for each uncertain coefficient $\Tilde{a}_{ij}$ as $\eta_{ij} = \dfrac{\Tilde{a}_{ij} - {a}_{ij}}{\hat{a}_{ij}}$ that represents the scaled perturbation of $\Tilde{a}_{ij}$ from its nominal value $a_{ij}$.

Bertsimas and Sim (2004) also introduced a parameter $\Gamma_{i} \in [0,|J_{i}|]$ as the budget of uncertainty for each constraint \textit{i}, where $|J_{i}|$ denotes the number of elements of set $J_{i}$. In fact, $\Gamma_{i}$ is the maximum number of parameters that can really deviate from their nominal values for each constraint \textit{i}. The parameter $\Gamma_{i}$ that bounds the total scaled deviation of uncertain parameters as $\sum_{j \in J_{i}} |\eta_{ij}| \leq \Gamma_{i}$ adjusts the robustness of the proposed method against the level of solution conservatism. In particular, $\Gamma_{i}=0$ represents the nominal or deterministic formulation, whereas $\Gamma_{i}=|\eta_{ij}|$ relates to the worst-case formulation in which all uncertain parameters are fixed at their worst-case values from the uncertainty set. However, decision maker can make a trade-off between the protection level of constraint \textit{i} and the degree of conservatism of the solution if $\Gamma_{i} \in (0,|J_{i}|)$. Therefore, the budget of uncertainty $\Gamma_{i}$ that is an input to the robust optimization model can specify how risk averse the decision-maker is. 

Bertsimas and Sim (2004) proposed a nonlinear programming model as follows, which is equivalent to the the uncertain model (1):
\begin{flalign} \nonumber 
Min_{} \quad &\sum_{j}^{} c_{j}x_{j}\\ \nonumber 
s.t. \quad &\sum_{j}^{} a_{ij}x_{j} + \underset{\Omega}{max}\{\sum_{j \in S_{i}}^{} \hat{a}_{ij}x_{j} + (\Gamma_{i} - \left \lfloor \Gamma_{i} \right \rfloor)\hat{a}_{it_{i}}x_{j}\} \leq b_{i}; \qquad\forall i\\
&x_{j}\geq 0; \qquad\forall j &&
\end{flalign}
where $\Omega=\{S_{i}\cup\{t_{i}\}|S_{i}\subseteq J_{i}, S_{i}=\left \lfloor \Gamma_{i} \right \rfloor, t_{i} \in J_{i} \setminus S_{i} \}$ is defined as the uncertainty set. For a given optimal solution $x^{*}$ of problem (2), Bertsimas and Sim (2004) demonstrated that the protection function for constraint \textit{i} against uncertainty, which is $\beta_{i}(x^{*},\Gamma_{i})=\underset{\Omega}{max}\{\sum_{j \in S_{i}}^{} \hat{a}_{ij}x_{j} + (\Gamma_{i} - \left \lfloor \Gamma_{i} \right \rfloor)\hat{a}_{it_{i}}x_{j}\}$ can be formulated as the following linear programming problem:
\begin{flalign} \nonumber 
\beta_{i}(x^{*},\Gamma_{i}) = Max_{} \quad &\sum_{j \in J_{i}} \hat{a}_{ij}|x_{j}^{*}|\eta_{ij}\\ \nonumber 
s.t. \quad &\sum_{j \in J_{i}}^{} \eta_{ij} \leq \Gamma_{i}; \qquad\forall i\\
&0 \leq \eta_{ij}\leq 1; \qquad\forall i,j &&
\end{flalign}
According to the theory of strong duality, since problem (3) is always feasible and bounded for all $\Gamma_{i} \in [0,|J_{i}|]$, its dual problem is feasible and bounded as well. Therefore, replacing the dual problem of problem (3) into (2), Bertsimas and Sim (2004) derived the robust formulation of the uncertain linear programming problem (1) as follows: 
\begin{flalign} \nonumber 
Min_{} \quad &\sum_{j}^{} c_{j}x_{j}\\ \nonumber 
s.t. \quad &\sum_{j}^{} a_{ij}x_{j} + \lambda_{i}\Gamma_{i} + \sum_{j \in J_{i}}\mu_{ij} \leq b_{i}; \qquad\forall i\\ \nonumber
& \lambda_{i} + \mu_{ij}\geq \hat{a}_{ij}x_{j}; \qquad\forall i, j \in J_{i}\\ \nonumber 
& \mu_{ij} \geq 0; \qquad\forall i, j \in J_{i}\\ \nonumber
&\lambda_{i} \geq 0; \qquad\forall i\\
&x_{j}\geq 0; \qquad\forall j &&
\end{flalign}
where $\lambda_{i}$ and $\mu_{ij}$ are dual variables associated with the first and second constraints in programming problem (3), respectively.

if the number of uncertain coefficients in constraint \textit{i} that perturb from their respective nominal values is less that or equal to $\Gamma_{i}$, then the optimal solution from robust problem (4) will remain always feasible. However, if more than $\Gamma_{i}$ coefficients deviate from their nominal values, then the probability of violating constraint \textit{i} for an optimal solution $x_{j}^{*}$ is calculated as follows:
\begin{flalign} 
Pr(\sum_{j} \Tilde{a}_{ij}x_{j}^{*} < b_{i}) \leq 1-\varphi(\dfrac{\Gamma_{i} - 1}{\sqrt{|J_{i}|}}) &&
\end{flalign}
where $\varphi(.)$ is the cumulative distribution function of a standard normal random variable.

\section{Numerical examples}
In this section, we present an illustrative example to delineate how the proposed approach is used in practice. We assume there are two residence areas: $h=1,2,3$; two drop-off sites: $c=1,2$; three primary processors: $p=1,2,3$; and one secondary processor: $s=1$; in this example to process two e-waste products: $i=1,2$. Three materials are recovered at primary processors: $j=1,2,3$; which are sent to a secondary processor to be remanufactured into new recycled materials.

\begin{figure}[!ht]   
\centering
  \includegraphics[width=10cm]{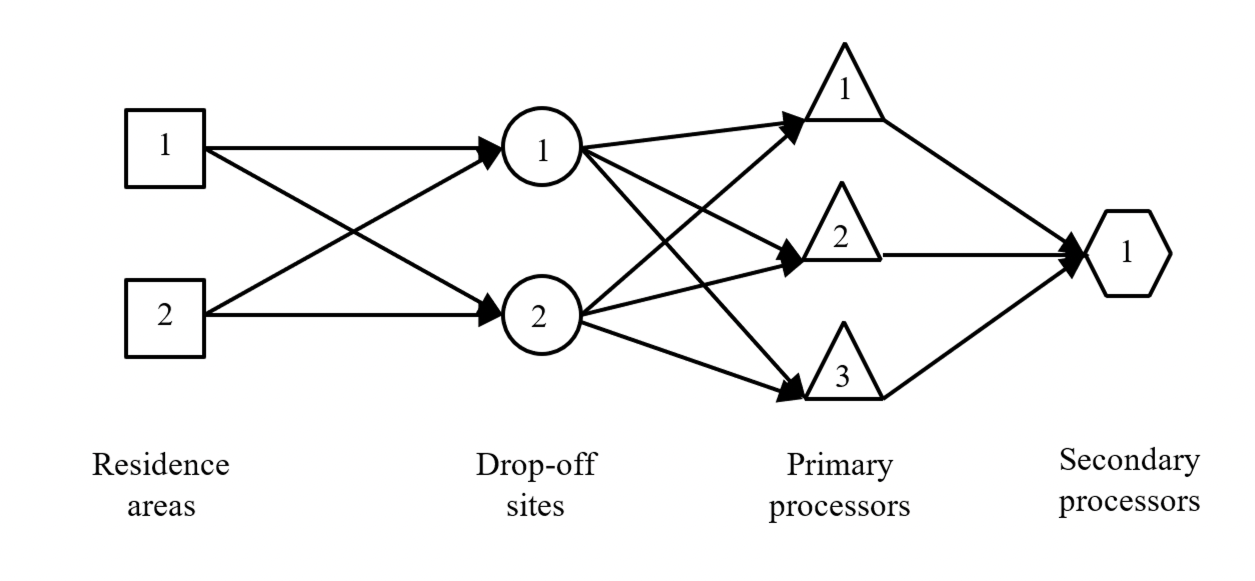}\\
  \caption{Illustration of the underlined logistics network.}
  \label{fig:numerical example}
\end{figure}

Figure~\ref{fig:numerical example} depicts the structure of the logistics network associated with the illustrative example. In order to study different aspects of the underlined problem and to signify the capability of the proposed approach to deal with them, we define several scenarios with different modifications on the configuration of the logistics network and input data, as follows:

\begin{enumerate} [i.]
    \item Base case scenario (BCS): In this scenario, we characterize the same product mix and quantity generated in each residence area and the same distances from drop-off sites 1 and 2 to all primary processors. Moreover, no capacity limits are forced to involved facilities and all input data related to the same types of facilities are assumed to be the same.
    \item Total capacity scenario (TCS): All input data for this scenario is the same as that for base case scenario. However, we put a binding total capacity limit on one of the primary processors, which is most utilized in the optimal solution of base case scenario, and reduce this capacity in several steps (80\% and 40\%) of throughput in base caseto illustrate how mass flows shifts and the network configuration changes.
    \item Product mix scenario (PMS): In this scenario, we change the products mix generated in each residence area. Specifically, we generate two instances of the data as follows: in instance 1, residence area 1 produces less e-waste than residence area 2. In instance 2, residence area 1 generates more of e-wastedevice 1 than residence area 2, while residence area 2 produces more of e-e-waste device 2 rather than residence area 1. All other parameters remain the same as those in base case scenario. 
\end{enumerate}

We implement the three scenarios on both system optimum and user optimum models and compare their obtained results. In addition, we assume 1050 kg of product 1 and 600 kg of product 2 are available in both residential areas. This product mix is used for the base case, total capacity, and distance vs. processing efficiency scenarios. However, for the different instances of product mix scenarios, we define the following product mix as given in Table~\ref{tab:PMSproductmix}.

\begin{table}[H]
\centering
\caption{Different product mix for instances of PMS}
\label{tab:PMSproductmix}
\begin{tabular}{lllll}
\hline
\multirow{2}{*}{} & \multicolumn{2}{l}{Instance   1} & \multicolumn{2}{l}{Instance 2} \\ \cline{2-5} 
 & Res.   Area 1 & Res. Area 2 & Res. Area 1 & Res. Area 2 \\ \hline
Product 1 & 600 & 1050 & 1050 & 600 \\
Product 2 & 600 & 1050 & 600 & 1050 \\ \hline
\end{tabular}
\end{table}

We assume the products are taken from residential areas to drop-off sites by personal cars of residents and between any two connected facilities in the remaining network by truck with appropriate capacity. It is recognized that not all user trips to drop off e-waste are made only for that purpose. We assume that the average fraction of trips that are dedicated (df) to e-waste is 0.5 for both residence areas. In addition, the average number of trips per participating household per year to return e-waste products to any drop-off sites is assumed to be 500 for both residence areas. The values of other parameters are given in \ref{app: input data}. The system optimum and user optimum models for the aforementioned scenarios are coded and solved in Python, utilizing a commercial  Gurobi solver on a dual-core 2.5 GHz computer with 8 GB RAM. In solving these different scenarios, we are particularly interested in investigating how system optimum model on different scenarios differ from the user optimum model on the corresponding scenarios in terms of cost and environmental impacts. While many potential environmental impacts could be computed, greenhouse gas (GHG) emissions, or global warming potential, were selected since GHG emissions and offsets occur at each step in the network, and because transportation emissions are significant.

\subsection{Results analysis}
\subsubsection{Transportation cost and emission}
\begin{figure}[H]   
\centering
  \includegraphics[width=15cm]{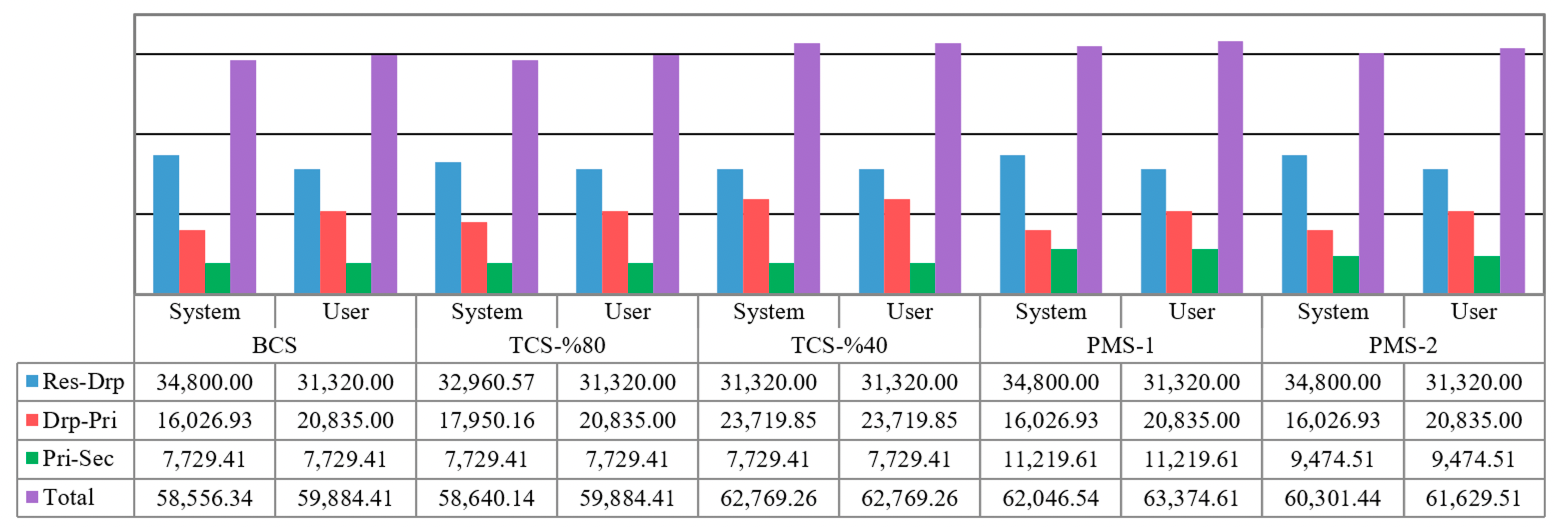}\\
  \caption{Transportation cost between connected facilities in the network.}
  \label{fig:transportationCost}
\end{figure}

\begin{figure}[H]   
\centering
  \includegraphics[width=15cm]{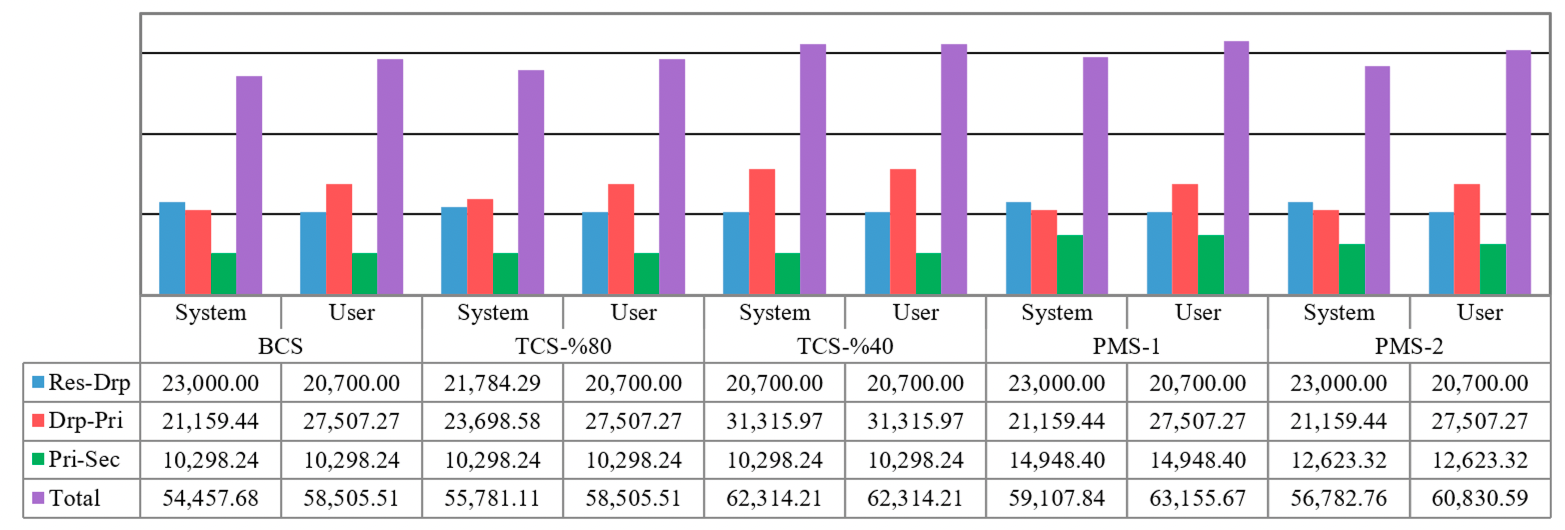}\\
  \caption{Transportation emission between connected facilities in the network.}
  \label{fig:transportationEmission}
\end{figure}

Two indicators in analyzing the results are the transportation cost and emissions between two connected facilities in the logistics network. As presented in Figures~\ref{fig:transportationCost}, \ref{fig:transportationEmission}, the transportation cost (emission) between residential areas to drop-off sites in system optimum solutions is greater than or equal to that in optimum user solutions. This observation is because the residential areas are assigned to drop-off sites in system optimum solutions in such a way that the total cost (emission) through the network is minimized, whereas this transportation cost (emission) is solely minimized by allotting residence areas to the closest drop-off sites in Model I of user optimum models. In addition, the residents are assumed to use their own personal vehicles to take the e-waste products to drop-off sites, rather than massively shipping the e-waste products and materials through the remaining parts of the network by trucks. Therefore, Figure 2 (Figure 3) shows that the transportation cost (emission) from residential areas to drop-off sites in all scenarios forms the major part of the total transportation cost (emission). Another observation that is common in all scenarios is that the transportation cost and emission from drop-off sites to primary processors in user optimum models are greater than or equal to those in system optimum models. This observation follows because, in user optimum models, the e-waste products are transported to drop-off sites that are only close to residential areas, but not necessarily to primary processors. 

In the total capacity scenarios (TCS), all parameters remain unchanged except the total capacity of the primary processors (third tier of the network ) decreases. In the base case scenario (regardless of the objective function), primary processor 3 is the most used. In fact, all products are transported from drop-off sites to this primary processor (capacity was high enough to process all e-waste). The total amount of available products in primary processor 3 is equal to 2.78 Mg. Hence, 80\% and 40\% of this quantity are regarded as the total capacity of primary processors in TCS-80\% and TCS-40\% scenarios, respectively. Decreasing the total capacity of primary processors requires transporting the extra e-waste products to farther primary processors based on their distances and processing costs. It leads to increased transportation costs and emissions from drop-off sites to primary processors in system optimum models of TCS-80\% and TCS-40\% rather than BSC. However, since the distance between any primary processor and the secondary processor is the same, transferring the extra e-waste products from primary processor 3 to other processors does not influence the transportation cost and emission from primary processors to the secondary processor in TCS-80\% and TCS-40\%. 

The unit costs and emissions for transporting two e-waste products from residence areas to drop-off sites and from drop-off sites to primary processors are the same. That is why changing the e-waste product mix in PMS-1 and PMS-2 does not change the transportation cost and emission from residence areas to drop-off sites and from drop-off sites to primary processors in comparison with BCS. However, since the amount of e-waste product 2 in PMS-1 and PMS-2 is greater than that in BCS and since more materials are recovered from these devices in primary processors, the transportation cost and emission from primary processors to the secondary processor in PMS-1 and PMS-2 are greater than those in BCS. 

\subsubsection{Processing cost and emission}
\begin{figure}[H]   
\centering
  \includegraphics[width=15cm]{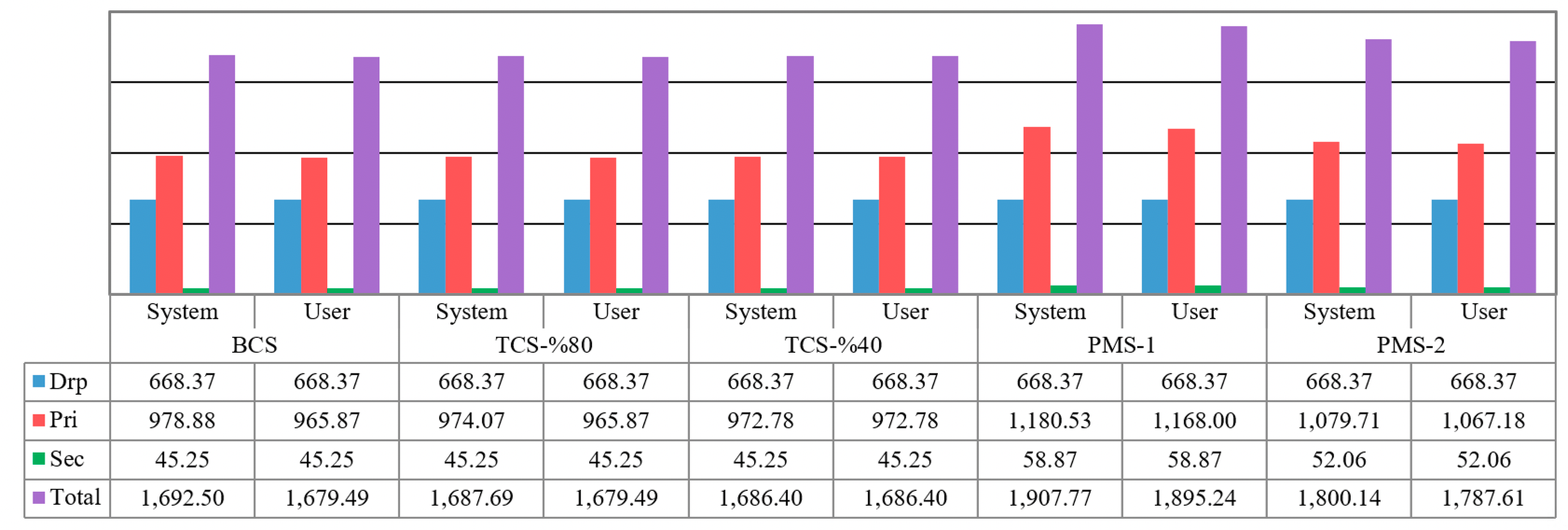}\\
  \caption{Processing cost for different facilities in the network.}
  \label{fig:process cost}
\end{figure}

Figure~\ref{fig:process cost} presents the total processing cost at different facilities of the logistics network. In this figure, the processing cost of drop-off sites is the same in all scenarios. It follows because all drop-off sites are assumed to have the same unit cost to process any type of e-waste. 

The total processing cost of primary processors in the user optimum model is less than or equal to that of the respective system optimum model for all scenarios. The reason for this observation is that the majority of e-waste products are processed by primary processor 3 in system optimum models of all scenarios because it is closer to drop-off site 1 in which almost all e-waste products are accumulated. However, in user optimum models, fewer e-waste products are processed by primary processor 3 and the remaining portion of them are transferred to primary processor 2, which, according to Table~\ref{tab:example process cost}, has smaller processing costs. A larger portion of e-waste products are transported to primary processor 2 in the latter case is because it is closer to drop-off site 2, to which residence area 2 is assigned in user optimum models of all scenarios. In addition, to verify why the processing cost of the secondary processor in user optimum and system optimum models of all scenarios is the same, we notice that, in each scenario, the amount of e-waste products in primary processors in user optimum model is the same as that in respective system optimum model. Since the fraction of material in an e-waste product is independent on the type of primary processors in this illustrative case, the amount of materials generated in all primary processors of user optimum model is equal to that of the system optimum model. Hence, we conclude the processing cost of the secondary processor in user optimum model is equal to that in system optimum model in all scenarios. 

The processing cost of primary processors in TCS-80\% and TCS-40\% are less than that in BCS. This observation follows because the extra e-waste products surpassing the total capacity of primary processor 3 are either completely transferred to primary processor 2 (as in TCS-80\%) or shipped to both processors 1 and 2 (as in TCS-40\%). According to Table B.2, the processing costs of processor 2 are less than those of processor 3, so it follows that the primary processors' cost in TCS-80\% is less than that in BCS. However, after depleting the capacity of primary processor 2 in TCS-40\%, the extra e-waste products are shipped to processor 1, which has slightly higher unit processing costs rather than the former two processors. Nevertheless, since only 557 kg of product 1 is processed by primary processor 1 in TCS-40\%, the total cost at primary processors in TCS-40\% is still less than that in BCS.

As previously stated, in analyzing the transportation costs, the amount of e-waste product 2 in PMS-1 and PMS-2 is greater than that in BCS. Since this e-waste product is more expensive to be processed by primary processors, their total processing cost in PMS-1 and PMS-2 is greater than that in BCS. Moreover, more materials are recovered from e-waste product 2 at the primary processors, which leads to a higher total processing cost of the secondary processor in PMS-1 and PMS-2.
 
\begin{figure}[H]   
\centering
  \includegraphics[width=15cm]{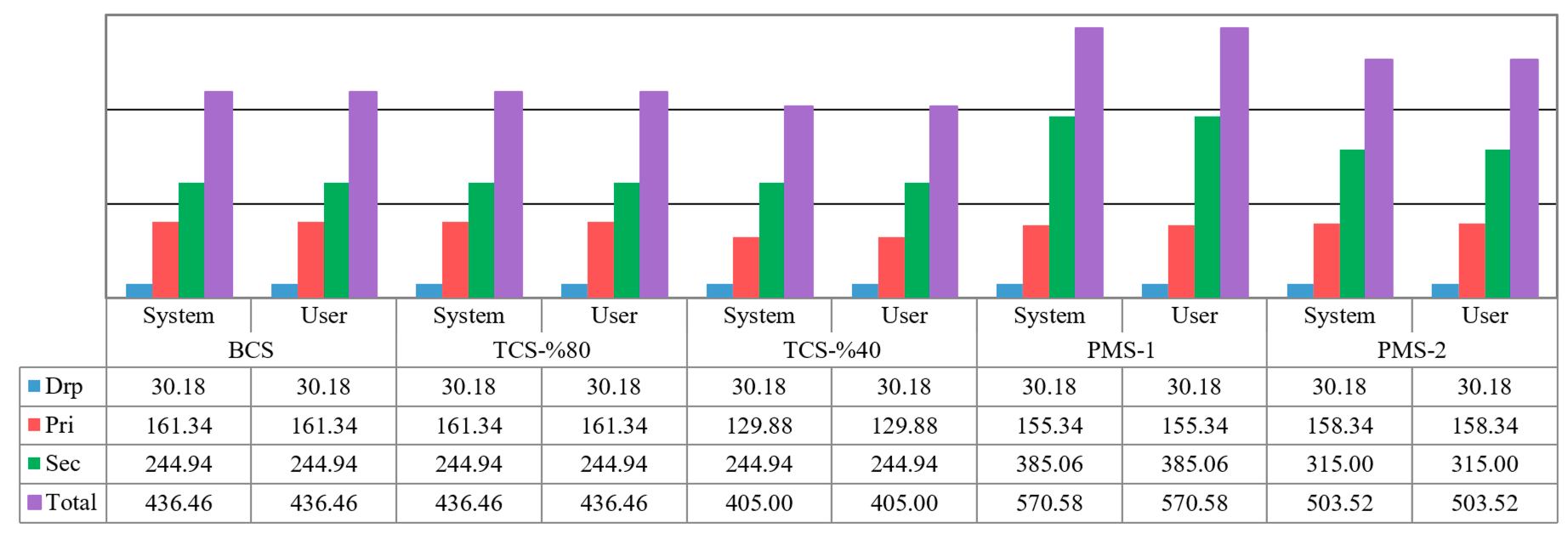}\\
  \caption{Processing emission for different facilities in the network.}
  \label{fig:process emission}
\end{figure}

Figure~\ref{fig:process emission} shows the processing emission in different facilities through the network.

Since all drop-off sites generate the same amount of emission to process the equivalent mass unit (i.e., kg) of any e-waste product (in this illustrative case), the processing emissions of drop-off sites is the same in both models of all scenarios. 
Moreover, the processing cost of primary processors in the user optimum models of all scenarios is the same that in the corresponding system optimum models. This is expected when product mix and device material composition are constant. To verify this observation, we note that residence areas 1 and 2 are assigned to drop-off sites 1 and 2, respectively, according to the criterion for selecting the closest drop-off site in all scenarios of the user optimum models. Once collected at drop-off sites, the e-waste products are transported from sites 1 and 2 to primary processors 3 and 2 respectively in all scenarios except TCS-40\%. In the user optimum model of the latter scenario (TCS-80\%), all primary processors need to operate because of their limited total capacity. 
We now examine the system optimum models. In system optimum models of BCS, PMS-1, and PMS-2, all e-waste products are transported to drop-off site 2, i.e. both residence areas are assigned to this drop-off site. Since, drop-off site 2 is close to primary processor 3, all e-waste products are then shipped to the latter primary processor. As previously discussed, the processing emission of primary processors in user optimum models of BCS, PMS-1, and PMS-2 should be the same as that in system optimum models. Furthermore, in system optimum model of TCS-80\%, not all e-waste products can be processed by primary processors 3 due to its total capacity limit. In this case, the extra amount of e-waste is transported to processor 2. Since, in both system optimum and user optimum models of TCS-80\%, all e-waste products are processed by primary processors 2 and 3 with the same processing emission, we obtain the same processing emission in primary processors for both models of TCS-80\%. Finally, in system optimum model of TCS-40\%, all primary processors should operate to process all e-waste products. In this case, a primary processor is assigned to the nearest drop-off site, i.e. primary processors 1 and 2 are assigned to drop-off site 2 and primary processors 2 and 3 to drop-off site 1. Since primary processors 2 and 3 are closer to drop-off sites, they are used up till reaching their full capacities. The extra amount of e-waste products that are related to product 1 from drop-off site 2 is sent to primary processor 1. However, this assignment of primary processors to drop-off sites and the mass flows among them happen identically in the respective user optimum model. Therefore, we conclude that the processing emission of primary processors in both systems optimum and user optimum models of TCS-40\% is the same. 

The same logic applies to the processing cost in PMS-1 and PMS-2 to verify the processing emission of the secondary processor in both system optimum and user optimum models of all scenarios is the same. 
In the system optimum model of TCS-80\%, because of the capacity limit of primary processor 3, some e-waste products are transported to the primary processor 2. However, since these two primary processors have the same unit processing emission, the processing emission of the system optimum model of TCS-80\% is the same as that of BCS. Furthermore, after depleting primary processors 2 and 3 in the system optimum model of TCS-40\%, the remaining-waste products should be shipped to primary processor 1 since it has lower unit processing emissions. Therefore, the processing cost of primary processors in TCS-40\% is less than that in both BCS and TCS-80\%. 

PMS-1 and PMS-2 have smaller processing emission of primary processors than BCS since the amount of e-waste product 2 in the former two scenarios are greater than that in the latter scenario. However, the resale rate of e-waste product 2 in primary processors is greater than the first e-waste product. Hence, the total amount of e-waste products processed by primary processors in PMS-1 and PMS-2 is less than that in BCS. Due to the same processing emission for both e-waste products in primary processor 3, we conclude the processing emission of primary processors in PMS-1 and PMS-2 should be less than that in BCS.  
The processing emission of the secondary processor in PMS-1 and PMS-2 is greater than that in BCS because, according to the fraction of materials in each e-waste product, more materials are recovered from the increased amount of e-waste product 2 in the former two scenarios. Hence, the secondary processor needs to process more materials, which results in higher processing emissions.   

\subsubsection{Fixed cost, revenue/offset, and total cost/emission}
Table~\ref{tab:fixedCost} represents fixed cost, revenue from re-sold devices and materials, and total cost incurred through the logistics network. Moreover, Table~\ref{tab:EmissionOffset} gives emission offset and total emission generated. In these tables, the total cost and emissions in user optimum models of all scenarios are greater than or equal to those in the corresponding system optimum models. This observation follows because the user optimum models for all scenarios provide a sub-optimal solution to the respective system optimum scenarios. The sub-optimality condition itself is a natural observation because, as mentioned before, the user optimum model contains two separate models to study two distinct parts of the network, namely, from residential areas to drop-off sites and from drop-off sites to secondary processors. 

\begin{table}[H]
\centering
\caption{Fixed cost, revenue, and total cost through the logistics network.}
\label{tab:fixedCost}
\begin{tabular}{lllll}
\hline
Scenario & Model & Fixed cost & Revenue & Total cost \\ \hline \hline
\multirow{2}{*}{BCS} & Syst. Opt. & 200 & 2,471 & 57,978 \\
 & Use. Opt. & 400 & 2,471 & 59,493 \\ \hline
\multirow{2}{*}{TCS-80\%} & Syst.   Opt. & 400 & 2,471 & 58,257 \\
 & Use.   Opt. & 400 & 2,471 & 59,493 \\ \hline
\multirow{2}{*}{TCS-40\%} & Syst. Opt. & 500 & 2,471 & 62,485 \\
 & Use. Opt. & 500 & 2,471 & 62,485 \\ \hline
\multirow{2}{*}{PMS-1} & Syst.   Opt. & 200 & 3,115 & 61,040 \\
 & Use.   Opt. & 400 & 3,115 & 62,555 \\ \hline
\multirow{2}{*}{PMS-2} & Syst. Opt. & 200 & 2,793 & 59,509 \\
 & Use. Opt. & 400 & 2,793 & 61,024 \\ \hline
\end{tabular}
\end{table}

\begin{table}[H]
\centering
\caption{Emission offset and total emission through the logistics network.}
\label{tab:EmissionOffset}
\begin{tabular}{lllll}
\hline
Scenario & Model & Offset & Total emission \\ \hline \hline
\multirow{2}{*}{BCS} & Syst. Opt. & 4,481 & 50,413 \\ 
 & Use. Opt. & 4,481 & 54,461 \\ \hline
\multirow{2}{*}{TCS-80\%} & Syst.   Opt. & 4,481 & 51,737 \\
 & Use.   Opt. & 4,481 & 54,461 \\ \hline
\multirow{2}{*}{TCS-40\%} & Syst. Opt. & 4,481 & 58,238 \\
 & Use. Opt. & 4,481 & 58,238 \\ \hline
\multirow{2}{*}{PMS-1} & Syst.   Opt. & 6,080 & 53,598 \\
 & Use.   Opt. & 6,080 & 57,646 \\ \hline
\multirow{2}{*}{PMS-2} & Syst. Opt. & 5,280 & 52,006 \\
 & Use. Opt. & 5,280 & 56,054 \\ \hline
\end{tabular}
\end{table}

In addition, total cost and emission in both models of TCS-80\% and TCS-40\% are greater than or equal to those in both respective models of BCS because the models in total capacity scenarios are more constrained than those in BCS. 
In PMS-1 and PMS-2, we verified in prior sections that their transportation and processing costs and emissions are greater than the transportation and processing costs and emissions of BCS. That is why total cost and emission of the two former scenarios are greater than those in the latter scenario. 

Moreover, in Table~\ref{tab:fixedCost}, we observe that the fixed cost in both models of TCS-80\% and TCS-40\% are greater than or equal to that in both models of BCS. It follows because more primary processors are needed to operate if their total capacity of them decreases. However, since only the mix of e-waste products changes in PMS-1 and PMS-2, the same number of primary processors are required to work in these two scenarios, which verifies the same fixed cost between the current scenarios and BCS. In each scenario, it is observed that the fixed cost of the user optimum model is greater than or equal to that of the corresponding system optimum model. This is because the residents are assigned to the closest drop-off sites in a user optimum model. In other words, the closeness to drop-off sites is the only criterion based on which the residents are allotted to drop-off sites and no other cost parameter plays a role in this allotment. Hence, more drop-off sites are required to cover all residence areas, which results in higher fixed cost in a user optimum model rather than its respective system optimum model. 

Since the unit processing credit (offset) remains the same in drop-off sites and primary processors and the mix of e-waste products does not change in the first three scenarios, the generated revenue in these scenarios is the same. However, in PMS-1, residence areas generate more of product 2 than product 1. eDevice 2 is largely decomposed into material 2, which is resold at a higher price in the secondary processor. Therefore, the revenue generated in PMS-1 is greater than that in BCS. The amount of e-waste product 2 in PMS-2 is greater (less) than that in BCS (PMS-1). Consequently, the yielded revenue in PMS-2 is greater (less) than that in BCS (PMS-1). On the other hand, the emission offset of e-waste product 2 in primary processors is far higher than that of e-waste product 1. Therefore, the total emission offset of PMS-1 and PMS-2 is greater than that of BCS.

\section{Case Study: Washington State E-waste management system}
Various take-back legislations are being implemented across the US \citep{atasu2009efficient, atasu2012operations}. A major concern in evaluating these programs is naturally the availability of required data.To cope with this issue, we examine implementing the proposed programming models in the State of Washington (WA), where reliable sources of required data for electronic devices are provided by the Washington State Department of Ecology’s e-cycling program \citep{WA2022reducing}.

The Washington State Department of Ecology manages implementing the take-back legislation, whereas the Washington Materials Management and Funding Authority (WMMFA) is responsible to perform the program. According to the State's law, original equipment manufacturers (OEMs) are allowed to create their own take-back programs in order to meet the requirements set forth in the law. Meanwhile, to date, no such “Independent Plan” has been approved, and all OEMs in the State operate under the Standard Plan. Under this standard plan, electronic products that constitute the e-waste stream include desktop computers, laptop computers, cathode ray tube (CRT) televisions and monitors, and flat-screen televisions and monitors. Although many other products and devices are considered electronic goods, this set was chosen to be representative of the areas in the US where a take-back program is in place. 

\begin{figure}[H]   
\centering
  \includegraphics[width=15cm]{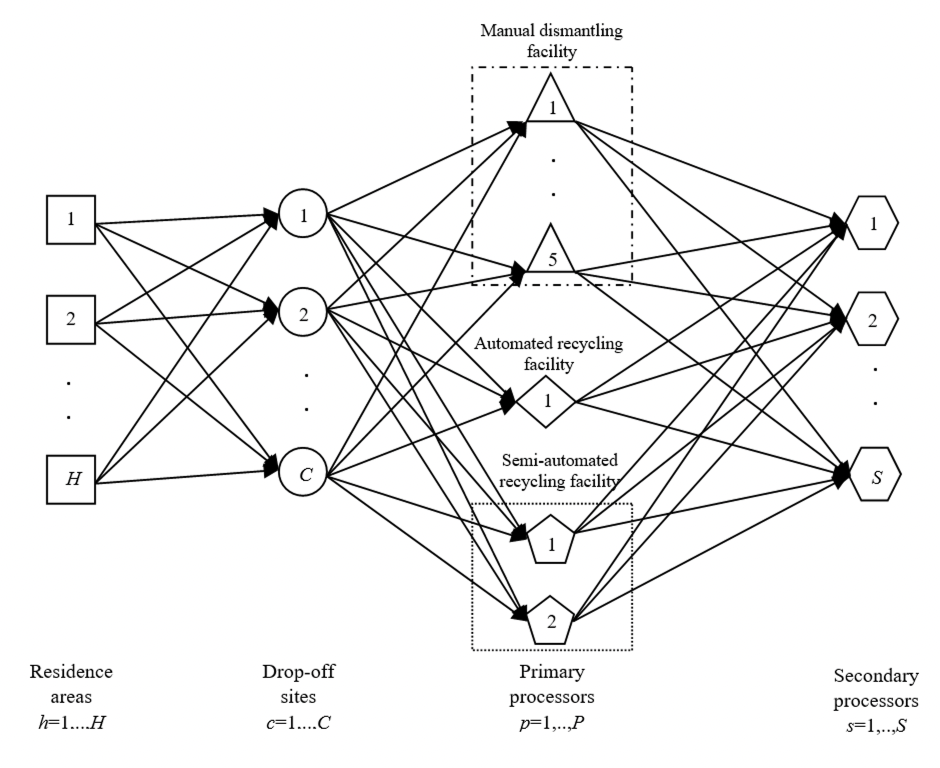}\\
  \caption{E-waste management system in Washington State.}
  \label{fig:RL network}
\end{figure}

Figure \ref{fig:RL network} shows the e-waste management system in Washington State, RL network begins with electronic devices that are no longer wanted by the current users and are going to be sent to end-of-life treatment. This includes e-waste generated by individuals, schools, small businesses, or small government operations (e.g., municipality offices) but excludes large private or public entities. While some devices may be functional and have a resale value, the system does not include asset management-type take-back legislation, which is typically a contractual arrangement involving the resale of devices anticipated to be functional at the end of their current use. Individuals and organizations bring or arrange for e-waste to be taken to 318 designated drop-off sites, where devices are sorted and re-sold if functional, or stored until they are shipped to one of 8 primary processing, or recycling, facilities. The primary processor may sort additional functional devices for resale, but the majority of them are dismantled through one of 5 available manual and/or an automated primary processors. Typically, batteries, mercury-containing devices (e.g., lamps), and circuit boards are manually removed. The remainder of the device may be further disassembled manually or may go through an automated process of size reduction and sorting. This includes shredding, magnet for ferrous metal recovery, eddy current separator for non-ferrous metal recovery, and some kind of optical- or density-based sort for plastic separation. The extent to which dismantling is automated is dependent on the technology and business model of each primary processing facility. After separating recoverable materials at the primary processor, each waste stream is sent to one of 28 appropriate secondary processing or remanufacturing facilities. Reprocessed materials, including aluminum, steel, plastics, and precious metals are output from the system to be sold as commodities. These materials offset the production of approximately equivalent virgin materials with respect to energy use and environmental emissions, as appropriate.

System and user optimum models presented in Section \ref{sec: model} are two general programming problems that can appropriately be implemented in various cases. However, there may be cases in practice in which we need to slightly modify these two general models to meet the requirements of the system under study. Particularly, according to take-back legislation in the Washington State, it is required locating at least one drop-off site in each county. Furthermore, a drop-off site should be placed at any city having a population of more than 10,000. As policy/legislation constraints, constraints \eqref{eq:Countypop1} and \eqref{eq:Countypop2} are enforced to imply together these requirements simultaneously, where $|k_u|$ is the cardinality of set $k_u$. Particularly, constraint \eqref{eq:Countypop1} is to place at least one drop-off site in each county and county \eqref{eq:Countypop2} denotes at least one drop-off site should be placed in each city with more than 10,000 populations. Certainly, later constraint is not taken into account for a county that does not have any such a city. 

\begin{flalign}
        \label{eq:Countypop1}  & \sum_{c \in \Bar{c}_{tu}}\sum_{ t \in c_u} X_{ctu} \geq Max\{1, |k_u|\} ; \quad \forall u \\ 
        \label{eq:Countypop2}  & \sum_{c \in \Bar{c}_{tu}} X_{ctu} \geq 1 ; \quad \forall u, t \in \{k_u \neq \varnothing \} &&
\end{flalign}

However, besides two general programming models presented in Section \ref{sec: model}, we examine two more set of instances of them in the Washington State case study, namely future projection and limited capacity. The future projection set of instances can be viewed as a sensitivity analysis of product mix and population growth, while the limited capacity instances can be regarded as a sensitivity analysis of capacities of primary processors.

\subsection{Data acquisition}
In this section, we will use various methods and data sources to estimate the values of our model parameters, including  cost components of our system (distance), and gridded residence areas to represent the spatial distribution of the population within our study region.

\textit{\textbf{Distance data.}}
Measuring the distance between consecutive facilities accurately seems a big challenge, because of the large number of facilities involved in the present case study.
There are various ways to do this, depending on the level of accuracy: The most accurate method is using an online application programming interface (API) mapping services to calculate the distance. These tools use routes and driving distances, so they provide accurate travel distances. There are a few reasons why calculating distance using this method takes a long time: 1. Server load: If servers of the API provider are experiencing a high volume of requests, it could take longer for your request to be processed, 2. Complexity of the request: calculating the distance between multiple locations or finding the optimal route between them could take longer for the API to process the request and return a response, and 3. Client-side processing: it could take longer to process the response and display the results on the user's device.

However, there is a close mathematical formula for calculating the distances between facilities on the earth's surface, called \textit{great circle} also known as \textit{Haversine formula} which appears to be an appropriate measure of distance, and faster due to the large number of facilities examined in this case study (Equation \eqref{eq:gedesic distance}). The great circle distances correspond to the shortest distance between two points on the surface (e.g., the earth) along the circle formed by the intersection of the surface and a plane passing through the center of the sphere. In this method, the elevations of points on the surface are usually ignored \citep{gade2010non}. 

\begin{equation} \label{eq:gedesic distance}
    d_{2,1} = 2 r sin^{-1}(\sqrt{sin^2(\frac{lat_2 - lat_1}{2}) + cos(lat_2) . cos(lat_1) . sin^2(\frac{long_2 - long_1}{2})})
\end{equation}

\begin{figure}[H]   
\centering
  \includegraphics[width=15cm]{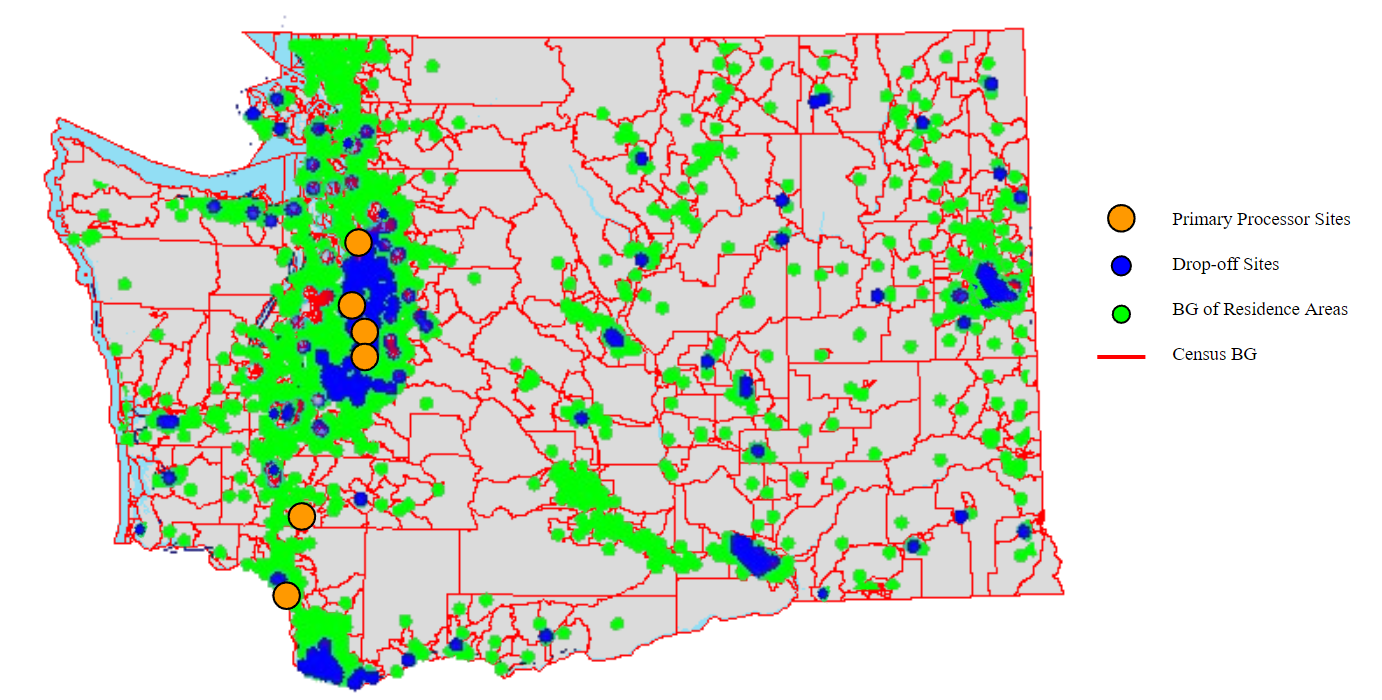}\\
  \caption{The illustration of residence areas (green), drop-off sites (blue), and primary processors (red) across the Washington State(Source of the base map: Census geographic files, office of financial management, Washington State, year 2020).}
  \label{fig:Census BG}
\end{figure}

\textit{\textbf{Gridding residence areas.}}
According to the 2020 census data of the Washington State, 7,705,281 persons were residing in this state in 2020. In order to derive the proper supply points for electronic e-wastes, we group this population based on the Block Groups (BGs), which are statistical divisions of census tracts generally defined to contain between 600 and 3,000 people and used to present data and control block numbering.  A block group consists of clusters of blocks within the same census tract that have the same first digit of their four-digit census block number.  
Grouping the residents of Washington State based on the Block Groups yields 4,762 blocks across the state, which are delineated by green points in Figure \ref{fig:Census BG}. The average population inhabiting and the average land area of these residential areas are equal to 1,618 and 12.04 square miles, respectively.

\subsection{Computational Results}
The system and user optimum models in conjunction with their sensitivity analysis-based instances are compared in this section using appropriate indicators. Particularly, we are interested in investigating how the system optimum model and its instances differ from the user optimum model and its corresponding instances. To accomplish this goal, all models and instances are coded and solved in Python, utilizing a commercial Gurobi solver a dual-core 2.5 GHz computer with 8 GB RAM.

One suitable indicator in analyzing the results is the transportation cost between two connected facilities in the logistics network. As illustrated in Figure (???), at the first glance, the total transportation cost in user optimum models in all instances is greater than that in the system optimum models. This observation follows because the user optimum models in all instances provide a sub-optimal solution to the system optimum instances. The sub-optimality condition itself is a natural observation because, as mentioned before, the user optimum model contains two separate models to study two distinct parts of the network, namely, from residential areas to drop-off sites and from drop-off sites to secondary processors. 

Meanwhile, the transportation cost between residential areas to drop-off sites in system optimum solutions is greater than that in user optimum solutions, because the residential areas are assigned to drop-off sites in system optimum solutions in such a way that the total cost through the network is minimized, whereas this transportation cost is solely minimized by allotting residence areas to drop-off sites in Model I of user optimum instances.
In addition, since the residents use their owned personal vehicles to take the wasted electronic devices to drop-off sites, rather than massively shipping the wasted devices and materials through the remaining parts of the network by trucks(!!), it turns out that the transportation cost from residential areas to drop-off sites in all instances forms the major part of the total transportation cost. It is at least 81.27\% and at most 94.09\% of the total transportation cost corresponding to the user optimum solution of 50\% limited capacity and system optimum solution of 2028 instances, respectively. 

Furthermore, as the total number of trips per year increases in 2018 rather than the baseline model, we observe that the transportation cost in 2018 for both system and user optimal solutions increases by 898,800 and 975,900 respectively. Since it is expected the total trips per year further increase in 2023 and 2028 due to population increase in the future, the transportation costs in these years for both system and user optimal solutions further increase in comparison with system and user optimum solutions in the present time.

%\subsection{Sensitivity Analysis}

\section{Conclusion}
In this paper, we proposed two different programming problems to measure economic and environmental effects of take-back legislation. According to the obtained results in both system optimum and user optimum models, among all stages of the life-cycle of e-waste products, the transportation phase between residence areas and drop-off sites has the greater contribution to the total cost and emission of the associated reverse logistics network. We also presented that how the optimum configuration of the reverse logistics network and mass flows for policy maker differ from those for the users in practice. 
Numerous parameters are involved in the proposed programming problem. Defining the precise values for them may be of great challenge in practice. Hence, rough estimations and imprecise values for them may be more applicable in some particular cases. Therefore, developing the proposed programming problems in this paper such that they are able to work with uncertain data can be a reasonable extension for future works. Moreover, the proposed mixed binary programming problems are intrinsically NP-hard problems. Thus, for big set of data, finding an optimum (or even almost optimum) solution for them may be costly or even impossible for practitioners. To deal with this challenge, one needs to develop efficient heuristic/metaheuristic algorithms that are able to reach a satisfactory solution in a reasonable amount of time.

%%%%%%%%%%%%%%%%%%%% Appendix
\newpage
\appendix
\section{Model parameters and decision variables}

\begin{table}[H]
\caption{Indexes and sets.}
\label{tab:index and set}
\begin{adjustbox}{width=1\textwidth}
\begin{tabularx}{\textwidth}{p{8cm} p{0.5cm} p{8cm}}
\hline
$i:$ Index of returned products e-waste;  $i \in \{1,   …, I\}$ & &
$s:$ Index of secondary processors; $s \in \{1, …, S\}$ \\

$h:$ Index of residence areas; $h \in \{1, …, H\}$ & & 
$j:$ Index of materials;  $j \in \{1, …, J\}$ \\

$c:$ Index of drop-off sites; $c \in \{1, …, C\}$ & &  
$l:$ Index of facility type; $l \in \{drp, pri, sec\}$ \\
$p:$ Index of primary processors; $p \in \{1, …, P\}$ & &
\\
\hline                                 
\end{tabularx}
\end{adjustbox}
\end{table}

\begin{table}[H]
\caption{Decision Variables.}
\label{tab:Decision Variables}
\begin{adjustbox}{width=1\textwidth}
\begin{tabularx}{\textwidth}{p{8cm} p{0.5cm} p{8cm}}
\hline
$RTD_{ihc}$: The fraction of trips for shipping $i$ from collection site $c$. 
& & $X_{c}$: Equal to one if collection site $c$ is opened; otherwise, zero.\\
 
$DTP_{icp}$: Amount of product $i$ shipped from collection site $c$ to primary processor $p$.
& & $Y_p$: Equal to one if primary processor $p$ is opened; otherwise, zero.\\ 

$PTS_{jpe}$: Amount of material $j$ shipped from primary processor $p$ to secondary processor to $s$.
& & $R_s$: Equal to one if secondary processor $s$ is opened; otherwise, zero.\\ 
\hline                                 
\end{tabularx}
\end{adjustbox}
\end{table}

\begin{table}[H]
\caption{Parameters.}
\label{tab:Parameters}
\begin{adjustbox}{width=1\textwidth}
\begin{tabularx}{\textwidth}{p{8cm} p{0.5cm} p{8cm}}
\hline

$q_{ji}$: The fraction of material $j$ in e-waste product $i$. 
& & $eff_{jp}^{pri}$: Separation efficiency of material $j$ at primary processor $p$.\\
 
$r_{ih}$: The total amount of e-waste product $i$ at residence area $h$.
& & $cap_{ic}^{drp}$: Maximum capacity of e-waste product $i$ drop-off site $c$\\ 

$pp_{h}$: The population of residence area $h$.
& & $cap_{ip}^{pri}$: Maximum capacity of wasted item $i$ at primary processor $p$. \\ 

$hs$: The average household size, persons per household.
& & $cap_{js}^{sec}$: Maximum capacity of material $j$ at secondary processor $s$. \\ 

$rq_{ic}$: The total amount of collected e-waste product $i$ in drop-off site $c$.
& & $pcr_{ip}^{drp}$: Unit processing credit of e-waste product $i$ in drop-off site $c$. \\ 

$df_{c}$: Dedicated fraction.
& & $pcr_{ip}^{pri}$: Unit processing credit of e-waste product $i$ at primary processor $p$. \\ 

$ty_h$: The average number of trips per participating household per year to return e-waste products to any drop-off sites.
& & $pcr_{js}^{sec}$: Unit processing credit of e-waste product $j$ at secondary processor $s$. \\ 

$nof_l$: Minimum number of facility type $l$ to be opened.
& & $pco_{ic}^{drp}$: Unit processing cost of wasted item $i$ in drop-off site $c$. \\ 

$re_{j}^{sec}$:  Fraction of material $j$ resold at secondary processor $j$.
& & $pco_{ip}^{pri}$: Unit processing cost of e-waste product $i$ at primary processor $p$. \\ 

$re_{j}^{l}$:  Fraction of e-waste product $i$ resold at facility type $l \neq sec$.
& & $pco_{js}^{sec}$: Unit processing cost of material $j$ at secondary processor $s$. \\

$env_{ic}^{drp}$: Unit $CO_2$ emission of e-waste product $i$ at drop-off site $c$.
& & $fcc_{c}$: Fixed cost of opening the drop-off site $c$. \\

$env_{ip}^{pri}$: Unit $CO_2$ emission of e-waste product $i$ at primary processor $p$.
& & $fcp_{p}$: Fixed cost of opening the primary processor $p$. \\

$env_{js}^{sec}$: Unit $CO_2$ emission of e-waste material $j$ at secondary processor $s$.
& & $fcs_{s}$: Fixed cost of opening the secondary processor $s$. \\

$ecr_{ic}^{drp}$: Unit $CO_2$ offset of e-waste product $i$ at drop-off site $c$.
& & $d_{hc}^{res-drp}$: Distance between residence area $h$ and drop-off site $c$. \\

$ecr_{ip}^{pri}$: Unit $CO_2$ offset of e-waste product $i$ at  primary processor $p$.
& & $d_{cp}^{drp-pri}$: Distance between drop-off site $c$ and primary processor $p$. \\

$ecr_{js}^{sec}$: Unit $CO_2$ offset of material $j$ at secondary processor $s$.
& & $d_{cp}^{pri-sec}$: Distance between primary processor $p$ and secondary processor $s$. \\

$env_{ihc}^{res-drp}$: Unit transportation $CO_2$ emission of e-waste product $i$ from residence area $h$ to drop-off site $c$.
&  & $lim_{ictu}^{drp}$: Minimum amount of e-waste product $i$ in drop-off site $c$. \\

$env_{icp}^{drp-pri}$: Unit transportation $CO_2$ emission of product $i$ from drop-off site $c$ to primary processor $p$.
&  & $lim_{ip}^{pri}$: Minimum capacity of e-waste product $i$ at primary processor $p$. \\

$env_{jps}^{pri-sec}$: Unit transportation $CO_2$ emission of material $j$ from primary processor $p$ to secondary processor $s$.
&  & $lim_{js}^{sec}$: Minimum capacity of material $j$ at secondary processor $s$. \\

$tco_{ihc}^{res-drp}$: Unit transportation cost of e-waste product $i$ from residence area $h$ to drop-off site $c$. 
&  & $tco_{ihc}^{drp-pri}$:  Unit transportation cost of e-waste product $i$ from drop-off site $c$ to primary processor $p$. \\

$tco_{jps}^{pri-sec}$: Unit transportation cost of material $j$ from primary processor $p$ to secondary processor $s$.
&  &  \\
\hline                                 
\end{tabularx}
\end{adjustbox}
\end{table}

\section{Input data for numerical example} 
\label{app: input data}

\begin{table}[H]
\caption{Distance ($km$), transportation cost ($\$/km$ for residence area to drop-off, $\$/kg-km$ for all others) and emission ($kg.CO_2/km$ for residence area to drop-off, $kg.CO_2/kg-km$ for all others) between relevant facilities.}
\label{tab:example distance}
\begin{adjustbox}{width=\columnwidth,center}
\begin{tabular}{lllllll}
\hline
 & Drop 1 & Drop 2 & Prim. 1 & Prim. 2 & Prim. 3 & Second.1 \\ \hline
Res. Area 1 & 100, 0.348, 0.23 & 150, 0.348, 0.23 & - & - & - & - \\
Res. Area 2 & 100, 0.348, 0.23 & 80, 0.348, 0.23 & - & - & - & - \\
Drop 1 & - & - & 150, 0.115, 0.152 & 100, 0.115, 0.152 & 50, 0.115, 0.152 & - \\
Drop 2 & - & - & 100, 0.115, 0.152 & 80, 0.115, 0.152 & 150, 0.115, 0.152 & - \\
Prim 1 & - & - & - & - & - & 3770, 3E-03, 3.6E-03 \\
Prim 2 & - & - & - & - & - & 3770, 3E-03, 3.6E-03 \\
Prim 2 & - & - & - & - & - & 3770, 3E-03, 3.6E-03 \\ \hline
\end{tabular}
\end{adjustbox}
\end{table}

\begin{table}[H]
\centering
\caption{Processing cost , processing revenue/credit ($\$/kg$) of product or material at each facility.}
\label{tab:example process cost}
\begin{tabular}{lllllll}
\hline
 & Drop 1 & Drop 2 & Prim. 1 & Prim. 2 & Prim. 3 & Second. 1 \\ \hline
Product 1 & 0.24, 3.12 & 0.24, 3.12 & 0.27, 0.04 & 0.25, 0.04 & 0.26, 0.04 & - \\
Product 2 & 0.24, 4.68 & 0.24, 4.68 & 0.62, 1.52 & 0.60, 1.52 & 0.61, 1.52 & - \\
Material 1 & - & - & - & - & - & 0.1, 0 \\
Material 2 & - & - & - & - & - & 0.05, 11.5 \\
Material 3 & - & - & - & - & - & 0.03, 2.3 \\ \hline
\end{tabular}
\end{table}

\begin{table}[H]
\centering
\caption{Processing emission, processing offset ($kg.CO2/kg$) of product or material at each facility.}
\label{tab:example process emmision}
\begin{tabular}{lllllll}
\hline
 & Drop 1 & Drop 2 & Prim. 1 & Prim. 2 & Prim. 3 & Second. 1 \\ \hline
Product 1 & 0.0108, 4.473 & 0.0108, 4.473 & 0.0044, 0.3465 & 0.062, 0.3465 & 0.062, 0.3465 & - \\
Product 2 & 0.0108, 10.85 & 0.0108, 10.85 & 0.0029, 6.3723 & 0.062, 6.3723 & 0.062, 6.3723 & - \\
Material 1 & - & - & - & - & - & 0.217, 1.639 \\
Material 2 & - & - & - & - & - & 0.403, 0.124 \\
Material 3 & - & - & - & - & - & 0.027, 3E-04 \\ \hline
\end{tabular}
\end{table}

\begin{table}[H]
\centering
\caption{Fraction of e-waste products and materials to resell at different facilities.}
\label{tab:example fraction resell}
\begin{tabular}{lllllll}
\hline
 & Drop 1 & Drop 2 & Prim. 1 & Prim. 2 & Prim. 3 & Second. 1 \\ \hline
Product 1 & 0.1561 & 0.1561 & 0.0194 & 0.0194 & 0.0194 & - \\
Product 2 & 0.1561 & 0.1561 & 0.1468 & 0.1468 & 0.1468 & - \\
Material 1 & - & - & - & - & - & 0.056 \\
Material 2 & - & - & - & - & - & 0.056 \\
Material 3 & - & - & - & - & - & 0.056 \\ \hline
\end{tabular}
\end{table}

\begin{table}[H]
\centering
\caption{Fraction of materials in each e-waste product.}
\label{tab:example fraction material}
\begin{tabular}{lllll}
\hline
 & Material 1 & Material 2 & Material 3 \\ \hline
Product 1 & 0.1038 & 0 & 0.0079 \\
Product 2 & 0.0345 & 0.62 & 0.0079 \\ \hline
\end{tabular}
\end{table}

%%%%%%%%%%%%%%%%%%% Bibliography
\newpage
\bibliographystyle{elsarticle-harv}\biboptions{authoryear}
\bibliography{references.bib}

\end{document}